\documentclass[twocolumn,showpacs,preprintnumbers,amsmath,amssymb,aps,prb]{revtex4} 
\usepackage{graphicx}
\usepackage{dcolumn}
\usepackage{bm} 
\usepackage{subfigure}  
\usepackage{natbib}  
\usepackage{multirow}
\usepackage{color}

\def\SBR{\xrightarrow{\mathrm{SBR}}}

\begin{document}

\title{Electron energy-loss and inelastic X-ray scattering cross
  sections \\ from time-dependent density-functional perturbation theory}

\author{Iurii Timrov and Nathalie Vast}
\affiliation{
  Laboratoire des Solides Irradi\'{e}s, \'Ecole Polytechnique - CEA - DSM - IRAMIS
  - CNRS UMR 7642,  91128 Palaiseau cedex, France
} 

\author{Ralph Gebauer}
\affiliation{
  ICTP -- The Abdus Salam International Centre for Theoretical
  Physics, Strada Costiera 11, 34014 Trieste, Italy
}

\author{Stefano Baroni}
\affiliation{
  Theory and Simulation of Materials (THEOS), \'Ecole Polytechnique
  F\'ed\'erale de Lausanne, 1015 Lausanne, Switzerland
} 
\affiliation{
  SISSA -- Scuola Internazionale Superiore di Studi Avanzati, Via
  Bonomea 265, 34136 Trieste, Italy\footnote{Permanent address}
}

\date{\today}

\begin{abstract}
  The Liouville-Lanczos approach to linear-response time-dependent
  density-functional theory is generalized so as to encompass electron
  energy-loss and inelastic X-ray scattering spectroscopies in
  periodic solids. The computation of virtual orbitals
  and the manipulation of large matrices are avoided by adopting a
  representation of response orbitals borrowed from (time-independent)
  density-functional perturbation theory and a suitable Lanczos
  recursion scheme. The latter allows the bulk of the numerical work
  to be performed at any given transferred momentum only once, for a
  whole extended frequency range. The numerical complexity of the
  method is thus greatly reduced, making the computation of the
  loss function over a wide frequency range at any given
  transferred momentum only slightly more expensive than a single
  standard ground-state calculation, and opening the way to
  computations for systems of unprecedented size and complexity.  Our
  method is validated on the paradigmatic examples of bulk silicon and
  aluminum, for which both experimental and theoretical results
  already exist in the literature.
\end{abstract}


\maketitle

\section{\label{sec:Introduction}Introduction}

Plasma oscillations in solids are possibly the simplest manifestation
of collective effects in condensed matter, and their understanding in
terms of \emph{plasmon} modes one of the earliest triumphs of quantum
many-body theory. \cite{Pines:1956,Nozieres:1959,Pines:1964} On the
experimental side, collective charge-density fluctuations can be
probed through electron energy-loss (EEL) or inelastic X-Ray
scattering (IXS) spectroscopies, two techniques that have been
steadily producing a wealth of data since the early 60s and 70s,
respectively.\cite{Egerton:1996,Schulke:2007} In the present day the
engineering of novel materials down to the nanometer scale makes it
possible to design devices where electromagnetic fields interact with
collective oscillations of structures of sub-wavelength size. The
strong dependence of plasmon dynamics on the size and shape of these
nanostructures holds the promise of an extraordinary control over the
optical response of the resulting devices, with applications to such
diverse fields as photovoltaics,\cite{Atwater:2011} proton beam
acceleration,\cite{Bartal:2011} or biosensing,\cite{Anker:2008} to
name but a few. This is plasmonics, \emph{i.e} photonics based on
collective electronic excitations in strongly heterogeneous systems,
where surface effects play a fundamental role. Plasma oscillations at
surfaces have recently aroused a renewed attention by themselves,
since it was shown that some metal surfaces unexpectedly exhibit
\emph{acoustic} plasmons.\cite{Silkin:2005, Diaconescu:2007,
  Pohl:2010, Vattuone:2012, Yan:2012, Vattuone:2013} These are
collective charge excitations localized at the surface, whose
frequency vanishes linearly with the wavevector, and are not damped by
the bulk electron-hole continuum.\cite{Silkin:2004, Pitarke:2004} It
is thought that these modes may offer the possibility of light
confinement at designated locations on the surface, with possible
applications in photonics and nano-optics.\cite{Pitarke:2007}

Most of the theoretical understanding of the optical response in
nano-plasmonic systems relies on a classical approach: the
nanostructure is usually described as an assembly of components, each
characterized by an effective macroscopic dielectric function, and
separated from the others by abrupt interfaces. The overall optical
response is then computed by solving Maxwell's equation for the
resulting heterogeneous system.\cite{Esteban:2012} When distances
between the nanoscale components are themselves nanometric, however,
quantum effects must be accounted for, and a fully quantum-mechanical
description is called for. 

Early quantum-mechanical approaches to the dynamics of charge-density
fluctuations\cite{Pines:1956,Nozieres:1959,Pines:1964} were based on
the random-phase approximation as applied to the jellium model
that, albeit exceedingly successful in simple metals and
semiconductors, is not suitable for more complex materials, nor can it
capture the fine, system-specific, features of even simple ones. The
effects of crystal inhomogeneities on plasmon resonances in
semiconductors (the so called \emph{local-field} effects) were first
addressed in the late 70s,\cite{Sturm:1978} using the empirical
pseudopotential method,\cite{Cohen:1966} along similar lines as
previously followed for the optical spectra.\cite{Louie:1975} In the
present day the method of choice for describing charge dynamics in
real materials (as opposed to simplified models, such as the jellium
one) is time-dependent (TD) density-functional theory
(DFT).\cite{Runge:1984,Gross:1996} Although some attempts to
investigate EEL and IXS spectra using many-body perturbation theory
have been
made,\cite{Caliebe:2000,Olevano:2001,Takada:2002,Arnaud:2005} the vast
majority of the studies existing to date relies on TDDFT, which in
fact has been successfully used to study plasmons in a number of
bulk\cite{Daling:1992,Engel:1992,Sturm:1992,Quong:1993,Fleszar:1995,Ehrnsperger:1997,Waidmann:2000,Vast:2002,Marinopoulos:2002,Marinopoulos:2003,Schone:2003,Dash:2004,Gurtubay:2004,Gurtubay:2005,Weissker:2006,Kramberger:2008,Alkauskas:2010,Huotari:2009,Weissker:2010,Cazzaniga:2011,Yan:2011} and surface\cite{Silkin:2005, Diaconescu:2007, Pohl:2010,
  Vattuone:2012, Yan:2012, Vattuone:2013} systems.

The conventional TDDFT approach to plasmon dynamics relies on the
calculation of the charge-density susceptibility, $\chi$ (or,
equivalently, inverse dielectric matrix, $\epsilon^{-1}$), starting
from the independent-electron susceptibility, $\chi_0$, via a
Dyson-like equation.\cite{Onida:2002} Although successful in
(relatively) simple systems that can be described by unit cells of
moderate size, this methodology can hardly be applied to more complex
systems, such as low-index or nano-structured surfaces, because of its
intrinsic numerical limitations. In particular: \emph{i}) the
calculation of $\chi_0$ requires the knowledge of a large number of
empty states, which is usually avoided in modern electronic-structure
methods; \emph{ii}) the solution of the Dyson-like equation requires
the manipulation (multiplication and inversion) of (very) large
matrices, and \emph{iii}) all the above calculations have to be
repeated independently for each value of the frequency to be sampled.

In this paper we introduce a new method, based on TD
density-functional perturbation theory
(DFPT),\cite{Walker:2006,Rocca:2008,Malcioglu:2011,Baroni:2012} that
allows to calculate EEL and IXS cross sections avoiding all the above
drawbacks, and thus lending itself to numerical simulations in complex
systems, potentially as large as several hundred independent atoms.
Although the new methodology is general in principle, our
implementation relies on the pseudopotential approximation, which
limits its applicability to valence (or shallow-core) loss
spectra. Inner-core loss spectra are currently addressed using
different methods, as explained \emph{e.g.}  in 
Refs.~\onlinecite{Soininen:2005, Joly:2001,Cabaret:2013}.  The salient features of
our method are: \emph{i})~the adoption of a representation from
time-independent DFPT\cite{Baroni:2001} allows to avoid the
calculation of Kohn-Sham (KS) virtual orbitals and of any large
susceptibility matrices ($\chi$ or $\chi_0$) altogether; and
\emph{ii}) thanks to the use of a Lanczos recursion scheme, the bulk
of the calculations can be performed only once for all the frequencies
simultaneously. The numerical complexity of the resulting algorithm is
comparable, for the \emph{whole} spectrum in a wide frequency range,
to that of a \emph{single} standard ground-state (or static response)
calculation.
 
The paper is organized as follows. In Sec.~\ref{sec:theory} we
describe our basic theoretical and algorithmic frameworks, including
the implementation of the newly proposed methodology for the response
of a periodic system to a monochromatic perturbation, relevant to the
calculation of EEL and IXS cross sections; in
Sec.~\ref{sec:applications} we benchmark our technique on the
prototypical examples of bulk silicon and aluminum, for which many
experimental and well established theoretical results already exist;
finally, our conclusions are presented in Sec.~\ref{sec:conclusions}.

\section{\label{sec:theory}  
  Theory and algorithms
}

Electron energy-loss spectroscopy probes the diffusion of a beam of
fast electrons through a solid. According to Van
Hove,\cite{VanHove:1954} the corresponding double-differential cross
section for inelastic scattering reads:\cite{Egerton:1996}
\begin{equation}
  \left( \frac{d^2\sigma}{d\Omega d\omega} \right)_\mathrm{EEL} =
  \left( \frac{4 \pi e^2}{Q^2} \right)^2 \frac{m^2}{4 \pi^2 \hbar^4}
  \frac{k_f}{k_i} \, S(\mathbf{Q},\omega) , 
  \label{eq:cross_section}
\end{equation}
where $-e$ and $m$ are the electron charge and mass, $\mathbf{k}_i$,
$\mathbf{k}_f$, and $\mathbf{Q} = \mathbf{k}_i - \mathbf{k}_f$ are the
incoming, outgoing, and transferred momenta, respectively, and
$S(\mathbf{Q},\omega)$ is the dynamic structure factor per unit
volume.

While EEL spectroscopy is not suitable for samples enclosed in
high-pressure cells, plasmon dynamics under pressure can be probed by
IXS spectroscopy.\cite{Mao:2001,Loa:2011} The double-differential
cross-section reads in this case:
\begin{equation}
  \left( \frac{d^2\sigma}{d\Omega d\omega} \right)_\mathrm{IXS} =
  \left( \frac{e^2}{m c^2} \right)^2 (\mathbf{e}_i \cdot
  \mathbf{e}_f)^2 \, \frac{\omega_f}{\omega_i} \, S(\mathbf{Q},\omega)
  ,  
  \label{eq:cross_section_IXS}
\end{equation}
where $\mathbf{e}_i$ and $\mathbf{e}_f$ are the incoming and scattered
photon polarization directions, and $\omega_i$ and $\omega_f$ are the
corresponding frequencies. According to the fluctuation-dissipation
theorem\cite{Pines:1966} $S(\mathbf{Q},\omega)$ is proportional to the
imaginary part of the charge-density susceptibility,
$\chi(\mathbf{Q},\mathbf{Q}; \omega)$:
\begin{equation}
  S(\mathbf{Q},\omega) = - \frac{\hbar}{\pi} \,
  \mathrm{Im} \, \chi(\mathbf{Q},\mathbf{Q}; \omega) . 
  \label{eq:fluct_dissip_theorem}
\end{equation}

In periodic solids the transferred momentum can be split into a
component in the first Brillouin zone, $\mathbf{q} $, and a
reciprocal-lattice vector, $\mathbf{G}$, as $\mathbf{Q} =
\mathbf{q}+\mathbf{G}$, and $\chi$ is often expressed in terms of the
inverse dielectric matrix, defined as:\cite{Martin:2004,Car:1981}
\begin{equation}
  \epsilon^{-1}_{\mathbf{G}\mathbf{G}'}(\mathbf{q},\omega) 
  = 
  \delta_{\mathbf{G}, \mathbf{G}'} + \frac{4\pi
    e^2}{|\mathbf{q}+\mathbf{G}|^2} \, 
  \chi(\mathbf{q}+\mathbf{G}, \mathbf{q}+\mathbf{G}';\omega) ,
  \label{eq:def_inv_microscopic_diel_tensor}
\end{equation}
where $\epsilon^{-1}_{\mathbf{G}\mathbf{G}'}(\mathbf{q},\omega) =
\epsilon^{-1}(\mathbf{Q},\mathbf{Q'};\omega)$. The function
$-\mathrm{Im}[\epsilon^{-1}(\mathbf{Q},\mathbf{Q};\omega)]$ is
usually referred to as the \emph{loss function}.

\subsection{
  Time-dependent density-functional perturbation  theory
}\label{sec:TDDFT}

In TDDFT electron dynamics is described by TD one-electron equations
for the occupied molecular orbitals. These TD KS equations
read:\cite{Martin:2004}
\begin{equation}
  i \, \frac{\partial \varphi_v(\mathbf{r},t)}{\partial t} =
  \hat{H}_{\mathrm{KS}}(t) \, \varphi_v(\mathbf{r},t), 
  \label{eq:TD-KS_equation}
\end{equation}
where $\varphi_v(\mathbf{r},t)$ and $\hat{H}_{\mathrm{KS}}(t)$ are the TD KS orbitals and Hamiltonian (quantum mechanical operators are
indicated with a caret), respectively, the index $v$ spans the $N_v$
occupied (\emph{valence}) states, and atomic units ($e=m=\hbar=1)$ are
used henceforth. The KS Hamiltonian reads:
\begin{equation}
  \hat{H}_{\mathrm{KS}}(t) = -\frac{1}{2} \nabla^2 +
  V_{ext}(\mathbf{r},t) + V_{\mathrm{HXC}}(\mathbf{r},t) , 
  \label{eq:KS_Hamiltonian}
\end{equation}
where $V_{ext}(\mathbf{r},t)$ and $V_{\mathrm{HXC}}(\mathbf{r},t)$ are
the external and Hartree-plus-exchange-correlation (HXC) potentials,
respectively. Let us assume
that the external potential can be split into a static term, plus a
small TD perturbation:
\begin{equation}
  V_{ext}(\mathbf{r},t) = V_{ext}^\circ(\mathbf{r}) +
  \lambda(t)V'_{ext}(\mathbf{r}) ,
  \label{eq:Vext}
\end{equation}
where $\lambda(t)$ is the TD strength of the perturbation. The total KS potential is perturbed accordingly:
$V'(\mathbf{r},t) = \lambda(t) V'_{ext}(\mathbf{r}) +
V'_{\mathrm{HXC}}(\mathbf{r},t)$, $V'_{\mathrm{HXC}}$ being the response HXC potential. The response of the KS orbitals is
defined as
\begin{equation}
  \varphi_v(\mathbf{r},t) = e^{-i \varepsilon_v t} \, \left (
    \varphi_v^\circ(\mathbf{r}) + \varphi'_v(\mathbf{r},t) \right) , 
\label{eq:orbital_response_funct}
\end{equation}
$\varphi_v^\circ(\mathbf{r})$ and $\varepsilon_v$ being the
unperturbed ground-state KS orbitals and energies, respectively. The
charge-density susceptibility is the response of the electron charge
density, which only depends on the projection of the response of the
valence KS orbitals onto the empty-state (conduction) manifold. The
Fourier transforms (indicated by tilde $``\tilde{\phantom{a}}"$
hereafter) of such projected response orbitals are obtained from
standard first-order perturbation theory via the linear systems:
\begin{equation}
  \left (\hat{H}^\circ - \varepsilon_v - \omega \right )
  \tilde{\varphi}'_v(\mathbf{r},\omega) = - \hat{P}_c
  \tilde{V}'(\mathbf{r},\omega) \varphi_v^\circ(\mathbf{r}) , 
  \label{eq:lin-resp_w_eq1}
\end{equation}
where $\hat P_c$ is the projector over the unperturbed
conduction-state manifold. Expressing the latter in terms of valence
orbitals ($\hat P_c+\hat P_v=1$) allows one to compute response KS
orbitals without making any reference to unoccupied states, much in
the same way as it is done in time-independent DFPT.\cite{Baroni:2001}
The solution of Eq.~(\ref{eq:lin-resp_w_eq1}) requires one to express
the total response potential, $\tilde
V'(\mathbf{r},\omega)$, in terms of its own solutions, through the
response charge density, which is the diagonal of the response density
matrix, $n'(\mathbf{r},t) =
  \rho'(\mathbf{r},\mathbf{r};t)$, whose Fourier transform is
defined as:
\begin{multline}
  \tilde{\rho}'(\mathbf{r},\mathbf{r}';\omega) = 2 \sum_{v=1}^{N_v} \bigl (
  \tilde{\varphi}^\prime_v(\mathbf{r},\omega) \,
  \varphi^{\circ\,*}_v(\mathbf{r}')  \\
  + \,   \varphi^\circ_v(\mathbf{r}) 
  \tilde{\varphi}^{\prime\,*}_v(\mathbf{r}',-\omega) \,
  \bigr ) ,
  \label{eq:charge-dens-response_2}
\end{multline}
where the factor two accounts for spin degeneracy in a non-polarized
system. Note that $ \tilde{n}'(\mathbf{r},\omega) =
\tilde{n}'^*(\mathbf{r},-\omega) $, as a consequence of the reality of
$n'(\mathbf{r},t)$. 
The equation for the complex conjugate of $\tilde{\varphi}^\prime_v(\mathbf{r},\omega)$
reads:
\begin{equation}
  \left (\hat{H}^\circ - \varepsilon_v + \omega \right)
  \tilde{\varphi}^{\prime\,*}_v(\mathbf{r},-\omega) = - \hat{P}_c
  \tilde{V}'(\mathbf{r},\omega) \varphi_v^{\circ\,*}(\mathbf{r}),
  \label{eq:lin-resp_w_eq2}
\end{equation}
where use has been made of the reality of the perturbing potential
($\tilde V'(\omega) = \tilde V'^*(-\omega)$). Equations~\eqref{eq:lin-resp_w_eq1} and \eqref{eq:lin-resp_w_eq2} describe the
\emph{resonant} and \emph{anti-resonant} contributions to
charge-density response, respectively. Their left-hand sides just
differ by the sign of the frequency, while, by using time-reversal symmetry
of the unperturbed system ($\varphi_v^{\circ\,*}=\varphi^\circ_v$) their
right-hand side can be made look alike.
The equations for the resonant and anti-resonant components of the
charge-density response are coupled by the HXC potential,
which is determined self-consistently by the density response itself,
through the relation:
\begin{equation}
  \tilde V'_{\mathrm{HXC}}(\mathbf{r},\omega) = \int
  \kappa(\mathbf{r},\mathbf{r}') \, \tilde{n}'(\mathbf{r}',\omega) \, d\mathbf{r}', 
\label{eq:V_HXC_2}
\end{equation}
where
\begin{equation} 
  \kappa(\mathbf{r},\mathbf{r}') = \frac{1}{|\mathbf{r-r'}|}+
  \frac{\delta V_{\mathrm{XC}}(\mathbf{r})}{\delta n(\mathbf{r}')} 
  \label{eq:HXC_kernel}
\end{equation}
is the HXC kernel, which we assume to be independent of frequency,
consistently with the adiabatic DFT approximation.\cite{Gross:1985}

The TD KS equations~(\ref{eq:TD-KS_equation}) can be equivalently
expressed in terms of a quantum Liouville equation for the
one-particle density matrix, $\hat{\rho}(t)$:\cite{Rocca:2008,Baroni:2012}
\begin{equation}
  i \, \frac{d\hat{\rho}(t)}{dt} = \left[ \hat{H}_{\mathrm{KS}}(t),
    \hat{\rho}(t) \right].
  \label{eq:Liouville_eq_1}
\end{equation}
Upon linearization and Fourier transformation,
Eq.~(\ref{eq:Liouville_eq_1}) takes the form:
\begin{equation}
  (\omega - \hat{\mathcal{L}}) \cdot \hat{\rho}'(\omega) =
  \tilde\lambda(\omega) [\hat{V}'_{ext}, \hat{\rho}^\circ ] ,
  \label{eq:Liouville_eq_FT_1_general}
\end{equation}
where $\hat{\rho}^\circ$ is the unperturbed density matrix and
$\hat{\mathcal{L}}$ is the Liouvillian super-operator, defined by the
relation:\cite{Rocca:2008,Baroni:2012}
\begin{equation}
  \hat{\mathcal{L}} \cdot \hat{\rho}' = [ \hat{H}^\circ, \hat{\rho}' ] +
  \left[ \hat V'_{\mathrm{HXC}}[\hat{\rho}'], \hat{\rho}^\circ \right] .
  \label{eq:Liouvillian_def}
\end{equation}
The response of an arbitrary one-electron Hermitian operator,
$\hat{A}$, to an external perturbation, $\hat V_{ext}$, is described
by the generalized susceptibility:
\begin{align}
  \chi_{AV}(\omega) & \equiv
  \frac{1}{\tilde\lambda(\omega)} \mathrm{Tr} \bigl
  ( \hat{A} \hat{\rho}'(\omega) \bigr )
  \label{eq:susceptibility_def_1} \\
  &=\bigl ( \hat{A}, (\omega - \hat{\mathcal{L}})^{-1} \cdot
  [\hat{V}'_{ext}, \hat{\rho}^0 ] \bigl ) ,
\label{eq:susceptibility_def_2}
\end{align}
where $(\cdot,\cdot)$ indicates a scalar product in an abstract
operator manifold.\cite{Malcioglu:2011}
Equation~\eqref{eq:susceptibility_def_2} states that, within TDDFT, the
most general susceptibility can be expressed as an off-diagonal
element of the resolvent of the Liouvillian.

\subsection{The Liouville-Lanczos algorithm}

The calculation of susceptibilities from
Eq. \eqref{eq:susceptibility_def_2} requires the explicit
representation of the response density matrix and of the Liouvillian
super-operator acting on it. The minimum dimension of such a
representation is $2\times N_v\times N_c$, where $N_c=N-N_v$ is the
number of virtual (conduction) orbitals and $N$ the dimension of
one-electron basis set.\cite{Timrov:Note:2013:2NcxNv} The inversion of
the Liouvillian appearing in Eq. \eqref{eq:susceptibility_def_2} is a
formidable task in typical large-scale plane-wave calculations, where
the number of occupied states can be as large as several hundreds to a
few thousands, and the number of virtual orbitals a hundred times as
large. The recursion method by Haydock, Heine, and Kelly
\cite{Bullet:1980} offers an elegant solution to a similar problem,
namely the calculation of a diagonal element of the resolvent of a
Hermitian matrix, in terms of a continued fraction, whose coefficients
are frequency-independent. The {\it Lanczos bi-orthogonalization
  algorithm},\cite{Saad:2003,Rocca:2008,Baroni:2012} allows one to generalize this
procedure to the calculation of \emph{off-diagonal} elements of the
resolvent of a \emph{non-Hermitian} matrix. The resulting numerical
workload for calculating the full spectrum in a whole wide frequency
range is comparable to that of a \emph{single} ground-state (or static
response) calculation. Other flavours of the Lanczos-type algorithm
can be found in Refs.~\onlinecite{Ankudinov:2002, Gruning:2011}.

\subsubsection{The Lanczos bi-orthogonalization algorithm}
\label{sec:Lanczos_method}

We want to calculate matrix elements such as:
\begin{equation}
  g(\omega) = \left( u, (\omega - L)^{-1} v \right) ,
  \label{eq:off_diag_matrix_element_general}
\end{equation}
where $L$ is a $P\times P$ non-Hermitian matrix, and $u$ and $v$ are
generic $P$-dimensional arrays. To this end we define two sets of
\emph{Lanczos vectors}, $\{v_j\}$ and $\{u_j\}$, through the recursive
relations:\cite{Saad:2003}
\begin{align}
  \beta_{j+1} \, v_{j+1} & = L \, v_j - \alpha_j \, v_j
  - \gamma_j \, v_{j-1} ,  \label{eq:Lanczos_chain_1} \\
  \gamma_{j+1} \, u_{j+1} & = L^\top \,
  u_j - \alpha_j \, u_j - \beta_j \, u_{j-1} , 
  \label{eq:Lanczos_chain_2}
\end{align}
where one defines $u_0=v_0=0$, $u_1=v_1=v$, and the $\alpha_j$,
$\beta_j$, and $\gamma_j$ \emph{Lanczos coefficients} are determined
by the \emph{bi-orthogonality} conditions $(u_j,v_j)=1$, and
$(u_{j-1},v_j)=(u_{j},v_{j-1})=0$. The set of vectors and
  coefficients generated through the recursion relations
  (\ref{eq:Lanczos_chain_1}-\ref{eq:Lanczos_chain_2}) is often referred to
  as a \emph{Lanczos chain}. 
The details of this algorithm are
reviewed \emph{e.g.} in Ref. [\onlinecite{Saad:2003}], and its
specialization to TDDFT is presented in
Refs. [\onlinecite{Rocca:2008,Baroni:2012}]. For the purposes of the
present paper, we limit ourselves to observe that the Lanczos vectors
thus generated have the property that they provide a tridiagonal
representation of the $L$ matrix. More specifically, if we define
the $P\times M$ matrices $^{M\!}U = \{ u_1, u_2, \ldots, u_M \}$ and
$^{M\!}V = \{ v_1, v_2, \ldots, v_M \}$ ($M$ being the number of
Lanczos iterations), one has:
\begin{equation}
\left(^{M\!}U\right)^\top L \,\, ^{M\!}V = \,^{M\!}T ,
\label{eq:L_to_T_tridiag}
\end{equation}
where $^{M\!}T$ is the tridiagonal matrix
\begin{equation}
^{M\!}T = \left(\begin{array}{ccccc}
\alpha_1 & \gamma_2  &    0     & \ldots   &  0       \\
\beta_2  & \alpha_2  & \gamma_3 &   0      &  \vdots  \\
0        & \beta_3   & \alpha_3 & \ddots   &  0       \\
\vdots   &    0      & \ddots   & \ddots   & \gamma_M \\
0        & \ldots    &    0     & \beta_M  & \alpha_M   
\end{array}\right) .
\label{eq:tridiagonal_matrix}
\end{equation}
In this Lanczos representation, the matrix element of
Eq.~\eqref{eq:off_diag_matrix_element_general} can be expressed as:\cite{Rocca:2008}
\begin{equation}
  g(\omega) \simeq \left ( ^{M\!}z , \left ( \omega \, ^{M\!}I  - \,
  ^{M\!}T \right )^{-1} \cdot \, ^{M\!}e_1 \right ) , 
\label{eq:resolvent_g}
\end{equation}
where
$^{M\!}e_1 = \{1,0,\ldots,0\}$ and $^{M\!} z$ is the $M$-dimensional
vector defined as:\cite{Rocca:2008,Baroni:2012} 
\begin{equation}
^{M\!} z =  \, \left( ^{M\!}V \right )^\top u .
\label{eq:zeta_coef_for_g}
\end{equation}
The right-hand side of Eq.~(\ref{eq:resolvent_g}) can be conveniently
computed by solving, for any given value of $\omega$, the equation:
\begin{equation}
\left( \omega \, ^{M\!}I - \, ^{M\!}T \right) {^{M\!}x} = \, ^{M\!}e_1 ,
\label{eq:eta_post_processing_for_g}
\end{equation}
and calculating the scalar product:
\begin{equation}
g(\omega) = \left ( {^{M\!} z} , {^{M\!} x} \right ) .
\label{eq:resolvent_Liouvillian_Lanczos_2_for_g}
\end{equation}
The vector $^{M\!} z$, Eq.~(\ref{eq:zeta_coef_for_g}), can be computed on
the fly during the Lanczos recursion, through the relation $z_j=\left
  (u,v_j\right )$. In practice, the procedure outlined above is
performed in two steps. In the first step, which is by far the most
time consuming, one generates the tridiagonal matrix $^{M\!}T$,
Eq.~(\ref{eq:tridiagonal_matrix}), and the vector $^{M\!}z$,
Eq.~(\ref{eq:zeta_coef_for_g}).  In the second step $g(\omega)$ is
calculated from Eq.~(\ref{eq:resolvent_Liouvillian_Lanczos_2_for_g})
upon the solution of Eq.~(\ref{eq:eta_post_processing_for_g}), for
different frequencies $\omega$. In practice, a small imaginary part
$\eta$ is added to the frequency argument, $\omega \rightarrow
\omega + i\eta$, so as to regularize the function
$g(\omega)$.\cite{Rocca:2008,Baroni:2012} Setting $\eta$ to a non-zero value
amounts to broadening each individual spectral line or,
alternatively, to convoluting the function $g(\omega)$ with a
Lorentzian. Because of the tridiagonal form and the small dimension of
the matrix $^{M\!}T$ (a few hundreds to a few thousands), the second
step is essentially gratis. Different responses to a same
perturbation can be computed simultaneously from a same Lanczos
recursion, by computing different $z$ vectors on the fly.

\subsubsection{The batch representation}
\label{sec:Batch_repr_general}

Equation~\eqref{eq:charge-dens-response_2} shows that the response
density matrix is uniquely determined by the two sets of functions $\{
\tilde{\varphi}^\prime_v(\mathbf{r},\omega) \}$ and $\{
\tilde{\varphi}^{\prime\,*}_v(\mathbf{r},-\omega) \}$. It is
convenient to consider a linear combination of these functions,
defined as:
\begin{align}
  q_v(\mathbf{r}) &= \frac{1}{2} \, \bigl (
  \tilde{\varphi}^\prime_v(\mathbf{r},\omega) +
  \tilde{\varphi}^{\prime\,*}_v(\mathbf{r},-\omega) \bigr )
  ,   \label{eq:batch_q} \\
  p_v(\mathbf{r}) &= \frac{1}{2} \, \bigl (
  \tilde{\varphi}^\prime_v(\mathbf{r},\omega) -
  \tilde{\varphi}^{\prime\,*}_v(\mathbf{r},-\omega) \bigr ) . 
  \label{eq:batch_p}
\end{align}
The two sets $\{q_v\}$ and $\{p_v\}$ are called respectively the {\it
  upper} and {\it lower} component of the \emph{standard batch
  representation} (SBR)\cite{Rocca:2008,Baroni:2012} of the response density
matrix \emph{super-vector}: $\hat\rho'\SBR \bigl\{ \{q_v\},\{p_v\}
\bigr\}$.\cite{Malcioglu:2011} The SBR of a Hermitian operator, $\hat
A$, has vanishing lower component, $\hat A \SBR \bigl\{ \{ \hat{P}_c \,
\hat A \, \varphi_v^\circ(\mathbf{r}) \}, 0 \bigr\}$, while that of its
commutator with the unperturbed density matrix [see
Eq.~(\ref{eq:Liouville_eq_FT_1_general})] has vanishing upper
component, $[\hat A, \hat{\rho}^\circ ] \SBR \bigl\{ 0, \{ \hat{P}_c \,
\hat A \, \varphi_v^\circ(\mathbf{r}) \} \bigr\}$. The SBR of the
Liouvillian super-operator has the block form:\cite{Rocca:2008,Baroni:2012}
\begin{equation}
  \hat{\mathcal{L}} =
  \left(
    \begin{array}{cc}
      0 & \hat{\mathcal{D}} \\ 
      \hat{\mathcal{D}} + \hat{\mathcal{K}} & 0
    \end{array}
  \right) ,
  \label{eq:Liouvillian_SBR}
\end{equation}
where the $\hat{\mathcal{D}}$ and $\hat{\mathcal{K}}$
super-operators are defined by their action on response batches,
\begin{align}
  \hat{\mathcal{D}} \{ q_v(\mathbf{r}) \} & = \bigl\{ (\hat{H}^\circ  -
  \varepsilon_v) q_v(\mathbf{r}) \bigr\},
  \label{eq:D_super-operator} \\
  \hat{\mathcal{K}} \{ q_v(\mathbf{r}) \} & = \Bigl\{
    4 \hat{P}_c \sum_{v'} 
     \int \kappa(\mathbf{r}, \mathbf{r}')
     \varphi^{\circ\,*}_{v'}(\mathbf{r}') q_{v'}(\mathbf{r}') 
     d\mathbf{r'} \, \varphi^\circ_v(\mathbf{r}) \Bigr\} \nonumber \\
     %
     & = \left \{  \hat{P}_c
    V'_\mathrm{HXC}(\mathbf{r}) \varphi^\circ_v(\mathbf{r}) \right \},
    \label{eq:K_super-operator'} 
\end{align}
$\kappa(\mathbf{r}, \mathbf{r}')$ is the HXC kernel
  of Eq.~(\ref{eq:HXC_kernel}), and $V'_\mathrm{HXC}$ is the HXC
  potential (see Eq.~\eqref{eq:V_HXC_2}) generated by the response charge
  density distribution whose SBR is (see
  Eq.~\eqref{eq:charge-dens-response_2}):
\begin{equation}
  n'(\mathbf{r}) = 4 \sum_v
  \varphi^{\circ\,*}_v(\mathbf{r}) \, q_v(\mathbf{r}) . 
  \label{eq:SBR_charge-density}
\end{equation}
According to the above equations,
operating with the Liouvillian on a test super-vector essentially
requires the calculation of the HXC potential response, its
application to each valence KS orbital, as well as the operation of
the unperturbed Hamiltonian onto twice the number of valence KS
states.

The starting super-vector of the Lanczos recursion is the right-hand
side of Eq. \eqref{eq:Liouville_eq_FT_1_general} whose SBR is:
\begin{equation}
  v_1 = u_1 \SBR  
  \left(\begin{array}{c}0 \\ 
      \{ \hat{P}_c \, \tilde{V}'_{ext}(\mathbf{r}) \, \varphi_v^\circ(\mathbf{r}) \}
    \end{array}\right) .
  \label{eq:Lanczos_starting_vect}
\end{equation}
Because of the special block structure of the Liouvillian,
Eq.~(\ref{eq:Liouvillian_SBR}), the SBR of odd Lanczos iterates have
vanishing upper components, whereas the even ones have vanishing lower
components. As a consequence, the number of response wavefunctions
onto which the unperturbed Hamiltonian must operate per Lanczos
iteration is halved. Also, the diagonal elements of the resulting
tridiagonal matrix (the $\alpha$ coefficients) are all vanishing.

\subsubsection{Lanczos-chain extrapolation}
\label{sec:extrapolation}

It was previously noted that the components of the vector $^{M\!}z$,
Eq.~(\ref{eq:zeta_coef_for_g}), decrease rather rapidly to zero,
whereas the $\beta_j$ (and $\gamma_j$) coefficients oscillate around
two distinct values for odd and even iterations, whose average is
approximatively equal to one half of the kinetic-energy cutoff (in a
plane-wave implementation), and whose difference is approximately
twice as large as the excitation gap in insulating or semiconducting
materials.\cite{Rocca:2008, Baroni:2012} This finding can be used to
speed up considerably the calculation by adopting a suitable
extrapolation technique. In practice, the Lanczos recursion is stoped
after $M_0$ iterations, such that the components of the $z$ array are
small enough. The dimension $M$ of the linear system,
Eq.~\eqref{eq:eta_post_processing_for_g}, is then set to a very large
(and to a large extent arbitrary) value. The $z$ components from
$M_0+1$ to $M$ are set to zero, whereas the corresponding $\beta$ and
$\gamma$ coefficients are set to the average of the values that have
been actually computed. The accuracy of the calculated spectrum is
then checked {\it a posteriori} with respect to the value of $M_0$. In
many applications it turns out that $M_0$ may vary from a few hundreds
up to a few thousands (depending on the plane-wave kinetic energy
cutoff), and $M$ is a (to a large extent arbitrary) number reaching up
to several thousands. As the solution of tridiagonal systems can be
performed very efficiently via standard factorization techniques, the
numerical overhead of this procedure is negligible. More on Lanczos
extrapolation can be found in Refs.~\onlinecite{Rocca:2008,
  Malcioglu:2011,Baroni:2012}. 

\bigskip\bigskip

\subsection{A Liouville-Lanczos approach to EEL and IXS spectroscopies
  in crystals} \label{sec:LL_approach_for_EELS_IXS}

In a periodic solid the unperturbed KS orbitals are
$\varphi^\circ_v(\mathbf{r}) = \varphi^\circ_{n,\mathbf{k}}(\mathbf{r})$,
where $\{v\} = \{n,\mathbf{k}\}$, $n$ is a band index and $\mathbf{k}$
a point in the Brillouin zone. These KS orbitals can be cast into
the Bloch form:
\begin{equation}
  \varphi^\circ_{n,\mathbf{k}}(\mathbf{r}) = e^{i\mathbf{k}\cdot\mathbf{r}}
  \, u^\circ_{n,\mathbf{k}}(\mathbf{r}) , \label{eq:Bloch_function}
\end{equation} 
where $u^\circ_{n,\mathbf{k}}(\mathbf{r})$ is the lattice-periodic
function. Similarly, the total perturbing potential can be conveniently decomposed into \emph{Bloch components}:
\begin{equation}
  \tilde{V}'(\mathbf{r},\omega) = 
  \sum_{\mathbf{q} } \mathrm{e}^{i\mathbf{q}\cdot\mathbf{r}}
  \, \tilde{v}'_\mathbf{q}(\mathbf{r},\omega) , 
  \label{eq:Vext_q_decomposition}
\end{equation}
where $\tilde{v}'_\mathbf{q}(\mathbf{r})$ is also lattice-periodic,
and the sum extends over the first Brillouin zone. A similar
decomposition can be applied to the external and HXC response
potentials. The response of each KS orbital can be correspondingly
expressed as a linear combination of the responses to each Bloch
component of the perturbing potential:
\begin{equation}
  \tilde\varphi'_{n\mathbf{k}}(\mathbf{r},\omega) = \sum_\mathbf{q}
  \mathrm{e}^{i(\mathbf{k}+\mathbf{q}) \cdot \mathbf{r}}
  \tilde u'_{n,\mathbf{k+q}}(\mathbf{r},\omega),
\end{equation}
where $\tilde u'_{n,\mathbf{k+q}}(\mathbf{r},\omega)$ is a lattice-periodic
response orbital that satisfies the equation:
\begin{multline}
  (\hat{H}^\circ_\mathbf{k+q} - \varepsilon_{n,\mathbf{k}} - \omega) \,
  \tilde{u}'_{n,\mathbf{k+q}}(\mathbf{r},\omega) = \\
  - \hat{P}_c^\mathbf{k+q} \, \tilde{v}'_\mathbf{q}(\mathbf{r},\omega)
  \, u_{n,\mathbf{k}}^\circ(\mathbf{r}).
  \label{eq:lin-resp_w_eq6}
\end{multline}
In Eq. \eqref{eq:lin-resp_w_eq6}, as well as in the rest of this
paper, quantum-mechanical operators bearing a wave-vector subscript
(such as $\hat{H}^\circ_\mathbf{k+q}$) or superscript (such as
$\hat{P}_c^\mathbf{k+q}$) are thought to operate on lattice-periodic
functions, and are defined in terms of their coordinate
representations as:
\begin{align}
  H^\circ(\mathbf{r},\mathbf{r}') &= \sum_{\mathbf{k}} \mathrm{e}^{i
    \mathbf{k}\cdot(\mathbf{r}-\mathbf{r}')}\, {H}^\circ_\mathbf{k}
  (\mathbf{r},\mathbf{r}')\,, \\
  P_c(\mathbf{r},\mathbf{r}') &= \sum_{\mathbf{k}} \mathrm{e}^{i
    \mathbf{k}\cdot(\mathbf{r}-\mathbf{r}')}
  P_c^{\mathbf{k}}(\mathbf{r},\mathbf{r}').
  \label{eq:Projector_on_empty_states_2}
\end{align}
The projector onto the conduction manifold in
Eq.~\eqref{eq:lin-resp_w_eq6} can be expressed in terms of the
periodic parts of the unperturbed Bloch functions as:
\begin{equation}
  {P}_c^{\mathbf{k}}(\mathbf{r,r'})  = \delta (\mathbf{r-r'}) - \sum_n
  u^\circ_{n,\mathbf{k}}(\mathbf{r}) \,
  u^{\circ\,*}_{n,\mathbf{k}}(\mathbf{r}') , 
\end{equation}
where the sum extends over all the occupied bands. A similar
decomposition into Bloch components holds for the response density
matrix, which reads in this case:
\begin{equation}
  \tilde{\rho}'(\mathbf{r},\mathbf{r}';\omega) =
  \sum_\mathbf{q}
  \mathrm{e}^{i\mathbf{q\cdot(r-r')}}
  \tilde\rho'_\mathbf{q}(\mathbf{r},\mathbf{r}',\omega), 
\end{equation}
where 
\begin{multline}
  \tilde\rho'_\mathbf{q}(\mathbf{r},\mathbf{r}';\omega) =
  2 \sum_{n,\mathbf{k}} \bigl (
  \tilde{u}^\prime_{n,\mathbf{k+q}}(\mathbf{r},\omega) \,
  u^{\circ\,*}_{n,\mathbf{k}}(\mathbf{r}') + \\ u^{\circ\,*}_{n,\mathbf{k}}(\mathbf{r}) \,
  \tilde{u}^{\prime\,*}_{n,\mathbf{-k-q}}(\mathbf{r}',-\omega) \,
  \bigr ) .
  \label{eq:rho_prime-periodic}
\end{multline}
The anti-resonant contribution to the density-matrix response in
Eq.~\eqref{eq:rho_prime-periodic} satisfies the equation:
\begin{multline}
  (\hat{H}^\circ_\mathbf{k+q} - \varepsilon_{n,\mathbf{k}} + \omega) \,
  \tilde{u}^{\prime\,*}_{n,\mathbf{-k-q}}(\mathbf{r},-\omega) = \\
  - \hat{P}_c^\mathbf{k+q} \, \tilde{v}'_\mathbf{q}(\mathbf{r},\omega)
  \, u_{n,\mathbf{k}}^\circ(\mathbf{r}) , 
  \label{eq:lin-resp_w_eq7}
\end{multline}
which can be obtained from Eq.~\eqref{eq:lin-resp_w_eq6} by complex
conjugation and simple manipulations deriving from time-reversal
invariance of the unperturbed system
($u^\circ_{n,\mathbf{k}}=u^{\circ\,*}_{n,\mathbf{-k}}$) and the reality of the
perturbing potential ($ \tilde v'_\mathbf{q}(\mathbf{r},\omega)=
\tilde v^{\prime \, *}_\mathbf{-q}(\mathbf{r},-\omega)$).

\subsubsection{Batch representation for periodic solids}

In analogy with Eq.~\eqref{eq:charge-dens-response_2},
Eq.~\eqref{eq:rho_prime-periodic} shows that the response density
matrix of a periodic solid to a perturbation of wave-vector
$\mathbf{q}$ is uniquely determined by the two sets of response
orbitals $\{ \tilde{u}'_{n,\mathbf{k+q}} (\mathbf{r}, \omega) \}$ and
$\{ \tilde{u}^{\prime\,*}_{n,\mathbf{-k-q}} (\mathbf{r}, - \omega)
\}$. Note that $n$ and $\mathbf{k}$ are running indices, whereas
$\mathbf{q}$ is fixed. The SBR can in this case be defined as:
\begin{align}
  q_{n,\mathbf{k+q}}(\mathbf{r}) &= \frac{1}{2} \, \bigl (
  \tilde{u}'_{n,\mathbf{k+q}}(\mathbf{r},\omega) +
  \tilde{u}^{\prime\,*}_{n,\mathbf{-k-q}}(\mathbf{r},-\omega) \bigr ) , 
  \\
  p_{n,\mathbf{k+q}}(\mathbf{r}) &= \frac{1}{2} \, \bigl (
  \tilde{u}'_{n,\mathbf{k+q}}(\mathbf{r},\omega) -
  \tilde{u}^{\prime\,*}_{n,\mathbf{-k-q}}(\mathbf{r},-\omega) \bigl ) . 
\end{align}
The two sets of response orbitals, $q_\mathbf{q}=\{q_{n,\mathbf{k+q}}
\} $ and $p_\mathbf{q}=\{p_{n,\mathbf{k+q}}\}$ satisfy the coupled set
of equations:
\begin{equation}
  \left(
    \begin{array}{cc}
      \omega & -\hat{\mathcal{D}}_\mathbf{q} \\
      -\hat{\mathcal{D}}_\mathbf{q}-\hat{\mathcal{K}}_\mathbf{q} &
      \omega
    \end{array}
  \right)
  \left(
    \begin{array}{c}
      q_\mathbf{q} \\
      p_\mathbf{q}
    \end{array}
  \right) =
  \left(
    \begin{array}{c}
      0 \\
      y_\mathbf{q}
    \end{array}
  \right) ,
  \label{eq:Liouvillian_eq_SBR_2}
\end{equation}
where $ y_\mathbf{q} = \{ \hat{P}_c^\mathbf{k+q} \tilde
v'_{ext,\mathbf{q}}(\mathbf{r}) u_{n,\mathbf{k}}^\circ(\mathbf{r}) \} $,
and $\hat{\mathcal{D}}_\mathbf{q}$ and $\hat{\mathcal{K}}_\mathbf{q}$
are the super-operators defined by the relations:
\begin{align}
  \hat{\mathcal{D}}_\mathbf{q} q_\mathbf{q} & = \left \{
    (\hat{H}^\circ_\mathbf{k+q} - \varepsilon_{n,\mathbf{k}}) \,
    q_{n,\mathbf{k+q}}(\mathbf{r}) \right \} \label{eq:D_super-operator_2} \\
  \hat{\mathcal{K}}_\mathbf{q} q_\mathbf{q} & = \left \{ 
    \hat{P}_c^\mathbf{k+q} \tilde
    v'_{\mathrm{HXC},\mathbf{q}}(\mathbf{r})
    u^\circ_{n,\mathbf{k}}(\mathbf{r}) \right \},
  \label{eq:K_super-operator_2}
\end{align}
and
\begin{equation}
  \tilde v'_{\mathrm{HXC},\mathbf{q}}(\mathbf{r}) =
  \int\kappa(\mathbf{r},\mathbf{r}') 
  n'_\mathbf{q}(\mathbf{r}') d\mathbf{r}',
  \label{eq:v'_{HXC,q}}
\end{equation}
is the HXC potential generated by the response charge density:
\begin{equation}
  n'_\mathbf{q}(\mathbf{r}) = 4 \sum_{n,\mathbf{k}}
  u^{\circ\,*}_{n,\mathbf{k}}(\mathbf{r}) \, q_{n,\mathbf{k+q}}(\mathbf{r})
  . \label{eq:SBR_charge-density_2}
\end{equation}
Equations~\eqref{eq:D_super-operator_2}, \eqref{eq:K_super-operator_2}, and
\eqref{eq:SBR_charge-density_2} are closely parallel to
Eqs.~\eqref{eq:D_super-operator}, \eqref{eq:K_super-operator'}, and
\eqref{eq:SBR_charge-density} of Sec. \ref{sec:Batch_repr_general}.

In practice, the sum over $\mathbf{k}$ points is limited to the
portion of the Brillouin zone that is irreducible with respect to the
small group of $\mathbf{q}$ and the resulting function symmetrized
accordingly, in close analogy with time-independent DFPT for
lattice-dynamical calculations.\cite{Baroni:2001} More about the
exploitation of crystal symmetry in the calculation of dynamical
charge-density susceptibilities can be found in Ref. \onlinecite{Timrov:2013}.

The $\chi(\mathbf{Q,Q};\omega)$ component of the charge-density
susceptibility is obtained from Eq.~\eqref{eq:susceptibility_def_2} as
the response of the $\mathbf{Q=q+G}$ Fourier component of the
charge-density operator, whose coordinate representation reads $\hat{n}(\mathbf{q+G}) \to \mathrm{e}^{i(\mathbf{q+G})\cdot \mathbf{r}}$,
to a monochromatic perturbation, $V'_{ext}(\mathbf{r}) =
\mathrm{e}^{i(\mathbf{q+G})\cdot \mathbf{r}}$. The SBR of the periodic
part of $\hat{n}(\mathbf{q+G})$ is $\bigl\{ \{ \hat 
P_c^\mathbf{k+q} \mathrm{e}^{i\mathbf{G\cdot r}} u^\circ_{n,\mathbf{k}}
\}, 0 \bigr \}$. The final expression for the susceptibility is:
\begin{equation}
  \chi(\mathbf{Q,Q};\omega) = \Bigl ( \bigl\{ \{ \hat
  P_c^\mathbf{k+q} \mathrm{e}^{i\mathbf{G\cdot r}} u^0_{n,\mathbf{k}}
  \}, 0 \bigr \}, \bigl \{q_\mathbf{q},p_\mathbf{q} \bigr \} \Bigl ),
  \label{eq:susceptibility_Q}
\end{equation}
where $ \bigl \{q_\mathbf{q},p_\mathbf{q} \bigr \} $ is the solution
of Eq. \eqref{eq:Liouvillian_eq_SBR_2}, obtained when the periodic
part of the external perturbing potential is $\tilde
v'_{ext,\mathbf{q}}(\mathbf{r})= \mathrm{e}^{i\mathbf{G\cdot r}}$.
 
In practice the susceptibility  in Eq.~\eqref{eq:susceptibility_Q} is computed
following the procedure outlined in Sec.~\ref{sec:Lanczos_method}
(see Eq.~\eqref{eq:resolvent_g}):
\begin{equation}
  \chi(\mathbf{Q},\mathbf{Q};\omega) \simeq \left(
  ^{M\!}z_\mathbf{q} , ( \omega \, ^{M\!}I - \, ^{M\!}T_\mathbf{q}
  )^{-1} \cdot \, ^{M\!}e_1 \right) , 
  \label{eq:resolvent_Liouvillian_Lanczos_q}
\end{equation}
where $^{M\!}T_\mathbf{q}$ is a tridiagonal matrix of dimension $M$ of
the form (\ref{eq:tridiagonal_matrix}), and $^{M\!}z_\mathbf{q} = (
z_{1,\mathbf{q}}, z_{2,\mathbf{q}}, \ldots, z_{M,\mathbf{q}})$ is an
$M$-dimensional array whose coefficients $z_{j,\mathbf{q}}$ are
defined as:
\begin{equation}
  z_{j,\mathbf{q}} = \left( \{ \{ \hat{P}_c^\mathbf{k+q}
    e^{i\mathbf{G}\cdot\mathbf{r}} u_{n,\mathbf{k}}^\circ(\mathbf{r}) \} ,
    0 \} , v_j \right) .
  \label{eq:zeta_coef_q}
\end{equation}

\subsubsection{Metals}

The Liouville-Lanczos approach for EEL and IXS spectroscopies can be
extended to metals by a suitable generalization of the smearing
technique introduced by de Gironcoli in the static case for
lattice-dynamical calculations.\cite{deGironcoli:1995,Baroni:2001} In the smearing approach, each KS energy level is broadened by a
smearing function $(1/\sigma) \, \tilde{\delta}(\varepsilon/\sigma)$,
which is an approximation to the Dirac $\delta$-function in the limit
of vanishing smearing width $\sigma$. The monochromatic $\mathbf{q}$ component of the charge-density response Eq.~(\ref{eq:SBR_charge-density_2}) can then be cast into the
form:
\begin{multline}
  n'_\mathbf{q}(\mathbf{r}) = 2 \sum_{n,\mathbf{k}}
  u^{\circ\,*}_{n,\mathbf{k}}(\mathbf{r}) \bigl(
    \overline{u}^\prime_{n,\mathbf{k+q}}(\mathbf{r},\omega)
  \bigr. \\
  \bigl. + \,
    \overline{u}^{\prime\,*}_{n,\mathbf{-k-q}}(\mathbf{r},-\omega) \,
  \bigr) , \label{eq:appendix_charge-dens-response_4}
\end{multline}
where the functions
$\overline{u}^\prime_{n,\mathbf{k+q}}(\mathbf{r},\omega)$ and
$\overline{u}^{\prime\,*}_{n,\mathbf{-k-q}}(\mathbf{r},-\omega)$
satisfy the equations:
\begin{align}
  (\hat{H}^\circ_{\mathbf{k+q}}  -\varepsilon_{n,\mathbf{k}} & - \omega ) \,
  \overline{u}'_{n,\mathbf{k+q}}(\mathbf{r}, \omega) = \nonumber\\ 
  & \quad - \bigl( \tilde{\theta}_{F;n,\mathbf{k}} -
  \hat{P}_{n,\mathbf{k}}^{\mathbf{k+q}} \bigr) \,
  \tilde{v}'_\mathbf{q}(\mathbf{r},\omega) \,
  u^\circ_{n,\mathbf{k}}(\mathbf{r}) ,  \\
  (\hat{H}^\circ_{\mathbf{k+q}} -\varepsilon_{n,\mathbf{k}} & + \omega ) \,
  \overline{u}^{\prime\,*}_{n,\mathbf{-k-q}}(\mathbf{r}, -\omega) =
  \nonumber \\
  & \quad - \bigl( \tilde{\theta}_{F;n,\mathbf{k}} -
  \hat{P}_{n,\mathbf{k}}^{\mathbf{k+q}} \bigr) \,
  \tilde{v}'_\mathbf{q}(\mathbf{r},\omega) \,
  u^\circ_{n,\mathbf{k}}(\mathbf{r}) 
\end{align}
(cf. with Eqs.~\eqref{eq:lin-resp_w_eq6} and \eqref{eq:lin-resp_w_eq7}), where
\begin{align}
  \hat{P}_{n,\mathbf{k}}^{\mathbf{k+q}} & = \sum_m^{occ}
  \beta_{n,\mathbf{k};m,\mathbf{k+q}} | u^\circ_{m,\mathbf{k+q}} \rangle
  \langle u^\circ_{m,\mathbf{k+q}} | , \\
  \beta_{n,\mathbf{k};m,\mathbf{k+q}} & =
  \tilde{\theta}_{F;n,\mathbf{k}}
  \tilde{\theta}_{n,\mathbf{k};m,\mathbf{k+q}}  \nonumber \\ 
  & \qquad \qquad \qquad + \, \tilde{\theta}_{F;m,\mathbf{k+q}}
  \tilde{\theta}_{m,\mathbf{k+q};n,\mathbf{k}} , 
\end{align}
$\tilde{\theta}_{F;n,\mathbf{k}} \equiv \tilde{\theta} [(\varepsilon_F
- \varepsilon_{n,\mathbf{k}})/\sigma]$ and
$\tilde{\theta}_{m,\mathbf{k+q};n,\mathbf{k}} \equiv
\tilde{\theta}[(\varepsilon_{m,\mathbf{k+q}} -
\varepsilon_{n,\mathbf{k}})/\sigma]$ being smooth approximations to the step-function, and $\varepsilon_F$ is the Fermi
energy. It can be easily verified that the coefficients
$\beta_{n,\mathbf{k};m,\mathbf{k+q}}$ vanish when any of its indices
refers to an unoccupied state. Therefore, the operator
$\hat{P}_{n,\mathbf{k}}^{\mathbf{k+q}}$ involves only a small number
of partially occupied bands, and the first-order variation of the
wavefunctions and of the charge density can be computed avoiding any
explicit reference to unoccupied states, much in the same way as for
insulating materials. More details about the Liouville-Lanczos
approach for metals can be found in Ref.~\onlinecite{Timrov:2013}.

\begin{figure*}[t] 
  \begin{center}
    \subfigure[]{\includegraphics[width=0.4\textwidth]{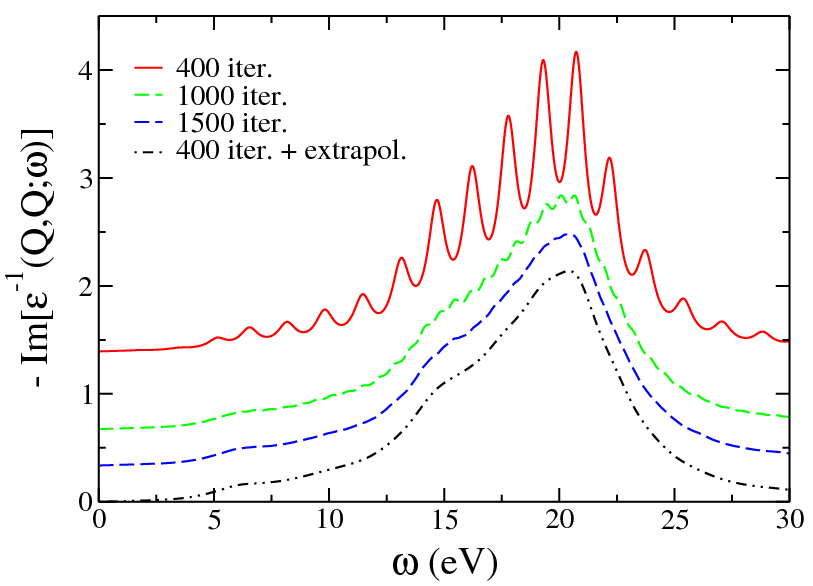}
      \label{fig:Si_conv_iter}}
    \hspace{0.3 cm}
    \subfigure[]{\includegraphics[width=0.4\textwidth]{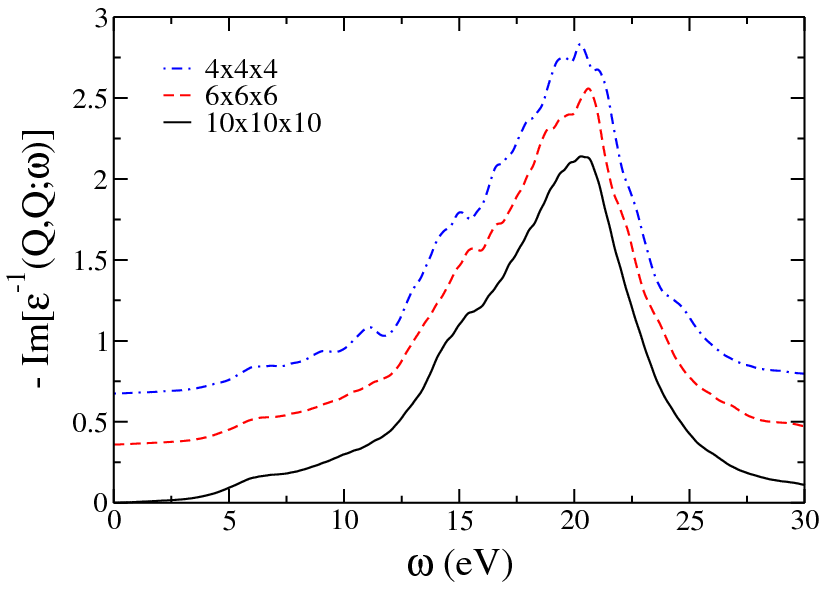}
      \label{fig:Si_conv_k}}
    \caption{Bulk Si: Loss function at
      $Q=0.53$ a.u. along [100].  
      (a)~Convergence with respect to the number of Lanczos iterations,
      and effect of the extrapolation technique, using a $10 \times 10
      \times 10$ Monkhorst-Pack $\mathbf{k}$ point mesh. (b)
      Convergence with respect to the size of the $\mathbf{k}$ point
      mesh, for 1500 Lanczos iterations. Both figures have been
      obtained with a Lorentzian broadening $\eta = 0.035$
      Ry. Curves have been shifted vertically for clarity.}
  \end{center}
\end{figure*}

\section{\label{sec:applications} Application to bulk Si and Al}

The technique described above has been implemented in the \textsc{Quantum ESPRESSO} suite
of computer codes,\cite{Giannozzi:2009} and is scheduled to be distributed in one of its
future releases. We now proceed to validate it by calculating the loss function in bulk
silicon and aluminum, for which several TDDFT studies exist, and whose spectra are known
to be accurately described within the adiabatic local density (LDA) and generalized
gradient (GGA) approximations (see, \emph{e.g.}, Refs.~\onlinecite{Weissker:2006,
  Weissker:2010} for Si, and \onlinecite{Cazzaniga:2011, Takada:2002} for Al).

All the calculations have been performed within the LDA approximation,
using the Perdew-Zunger parameterization of the electron-gas
data,\cite{Perdew:1981} norm-conserving pseudopotentials from the
\textsc{Quantum ESPRESSO} database\cite{Timrov:Note:2013:PP} and
plane-wave basis sets up to a kinetic-energy cutoff of 16 Ry. The
first Brillouin zone has been sampled with a Monkhorst-Pack (MP)
$\mathbf{k}$ point mesh, supplemented, in the case of Al, by the
Methfessel-Paxton smearing technique\cite{Methfessel:1989} with a
broadening parameter $\sigma=0.02$~Ry. The frequency argument of the
susceptibility has been assumed to have a small imaginary part,
$\eta$, thus resulting in a Lorentzian smearing of the spectra (see
Sec.~\ref{sec:Lanczos_method}). For both Si and Al we have used the
experimental lattice parameters (10.26 a.u.\cite{Neuberger:1971} and
7.60 a.u.,\cite{Wyckoff:1963} respectively), which is very close to
the theoretical one and resulting in no appreciable difference in the
computed spectra.

\subsection{Bulk silicon} 

Figure~\ref{fig:Si_conv_iter} shows the convergence of the loss spectrum of Si, as
calculated for a transferred momentum $Q=0.53$~a.u. along the [100] direction, as a
function of the number of Lanczos iterations. After 400 iterations the spectrum displays
spurious wiggles, which disappear by increasing the number of iterations up to 1500. Also
displayed are results obtained by the extrapolation procedure outlined at the end of
Sec. \ref{sec:Batch_repr_general}, performed with $M_0=400$ Lanczos iterations and
extrapolating the results up to a linear system of dimension $M=5000$. We see that the
numerical workload can be considerably reduced without any appreciable loss of accuracy.
In Fig.~\ref{fig:Si_conv_k} we show the convergence of the loss function with respect to
the $\mathbf{k}$ point sampling of the Brillouin zone. The $4 \times 4 \times 4$ MP
$\mathbf{k}$ point mesh is not dense enough to obtain a well-converged spectrum, due to
the presence of spurious wiggles, which disappear by increasing the size of the MP mesh up
to $10 \times 10 \times 10$.

\begin{figure*}[t]
\centering
\subfigure[]{\includegraphics[width=0.4\textwidth]{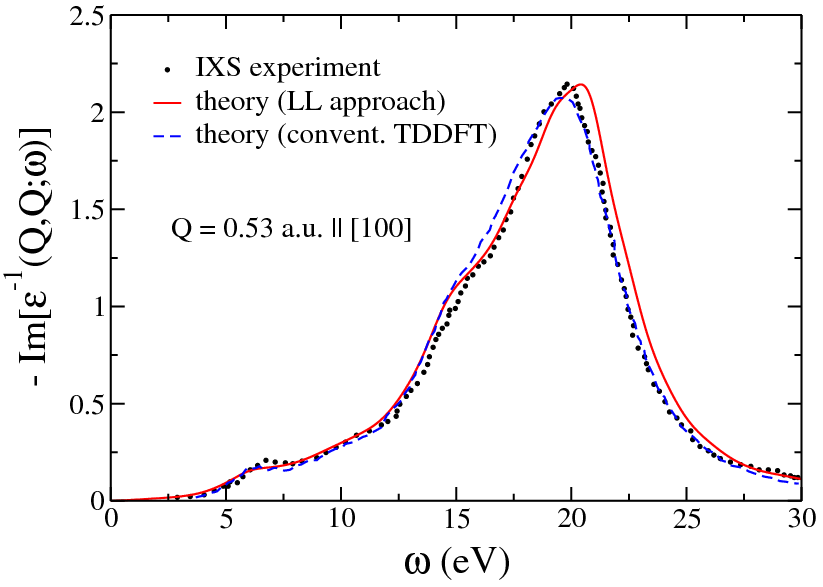}
\label{fig:Si_theor_vs_exp_1}}
\hspace{0.3 cm}
\subfigure[]{\includegraphics[width=0.4\textwidth]{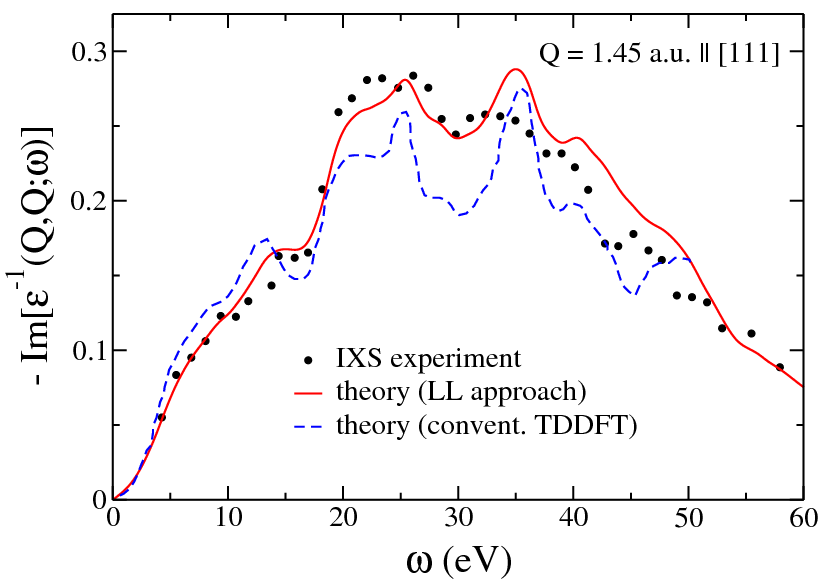}
\label{fig:Si_theor_vs_exp_2}}
\caption{Bulk Si: Comparison of the loss function calculated for two
  different values of the transferred momentum, $\mathbf{Q}$, using
  the Liouville-Lanczos (LL) approach, with
  experiment\cite{Weissker:2010} and with previous calculations.\cite{Weissker:2010} (a) $Q=0.53$ a.u. along [100], (b) $Q=1.45$~a.u. along [111]. LL data have been obtained using
  400 Lanczos iterations plus extrapolations. A $10 \times 10 \times
  10$ ($6\times 6\times 6$) MP $\mathbf{k}$ point mesh, and a Lorentzian broadening $\eta$ of
  0.035 (0.080) Ry have been used for the two cases, respectively.}
\label{fig:Si_theor_vs_exp}
\end{figure*}

In Fig.~\ref{fig:Si_theor_vs_exp} we compare our present results with those obtained from
the conventional approach based on the Dyson-like equation for the
susceptibility\cite{Onida:2002,Weissker:2010} and with experiment.\cite{Weissker:2010} The
agreement is excellent in both cases. All the salient features observed in the experiments
at small transferred momentum (panel (a)) are correctly predicted: the main plasmon peak
around 20 eV, a shoulder around 15 eV, and a weak peak around 6.5 eV. We attribute the
slight differences between the two theoretical spectra to the slightly different technical
details used in the two works.  In particular, the authors of
Ref.~\onlinecite{Weissker:2010} mimicked electron- and hole-lifetime effects with an
energy-dependent broadening, in contrast to the constant Lorentzian broadening, $\eta =
0.035$~Ry, used in our calculations. At larger momentum transfer (panel (b)) the
interaction of the plasmon with the electron-hole continuum broadens the
spectrum.\cite{Mahan:1990} The agreement with experiment,\cite{Timrov:Note:2013:IXS}
remarkable also in this case, is enhanced by increasing the Lorentzian broadening up to
$\eta =0.080$~Ry, which allowed us to reduce the size of the MP mesh down to $6 \times 6
\times 6$ without any appreciable loss of accuracy.

\begin{figure*}[t]
\begin{center}
\subfigure[]{\includegraphics[width=0.4\textwidth]{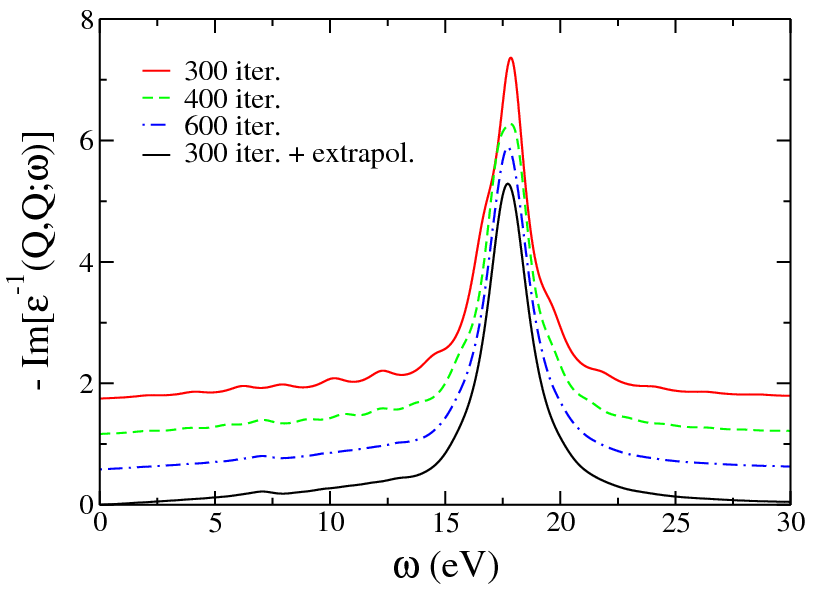}
\label{fig:Al_conv_iter}}
\hspace{0.3 cm}
\subfigure[]{\includegraphics[width=0.4\textwidth]{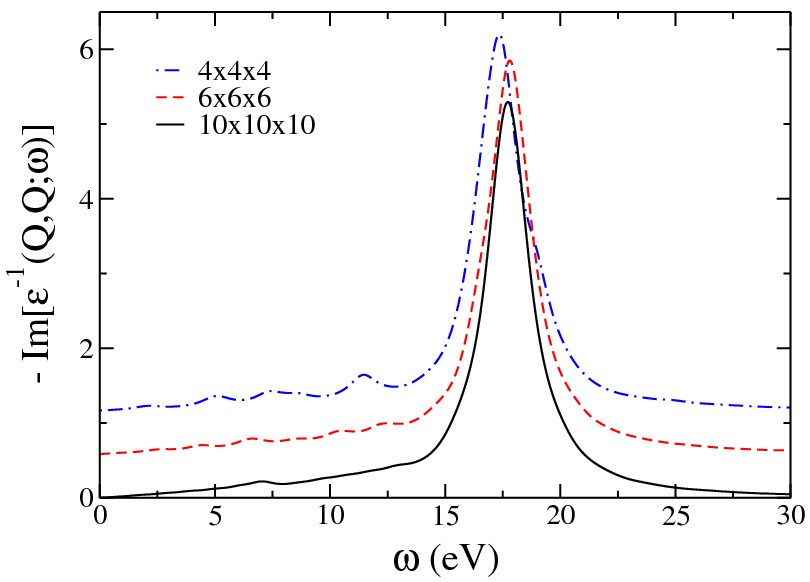}
\label{fig:Al_conv_k}} 
\caption{Bulk Al: Loss function calculated for
   $Q=0.513$ a.u. along [100].
  (a)~Convergence with
  respect to the number of Lanczos iterations, using a $10 \times 10
  \times 10$ MP $\mathbf{k}$ point mesh, and effect of the
  extrapolation technique. (b) Convergence with respect to the size of
  the $\mathbf{k}$ point mesh, using 600 Lanczos iterations. Both
  figures have been obtained with a Lorentzian broadening $\eta =
  0.056$ Ry. Curves have been shifted vertically for clarity.} 
\end{center}
\end{figure*} 

\subsection{Bulk aluminum} 

\begin{figure*}[t]
  \centering
  \subfigure[]{
    \includegraphics[width=0.4\textwidth]{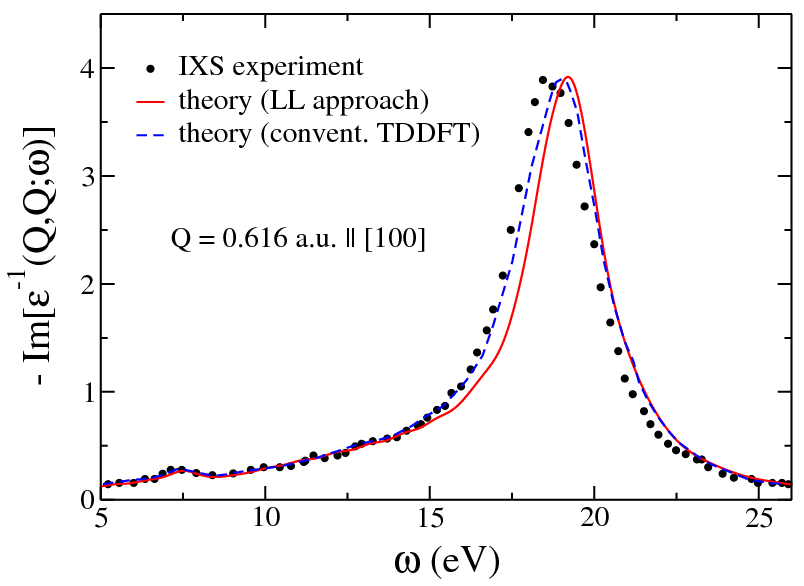}
    \label{fig:Al_theor_vs_exp_1}
  }
  \hspace{0.3 cm}
  \subfigure[]{
    \includegraphics[width=0.415\textwidth]{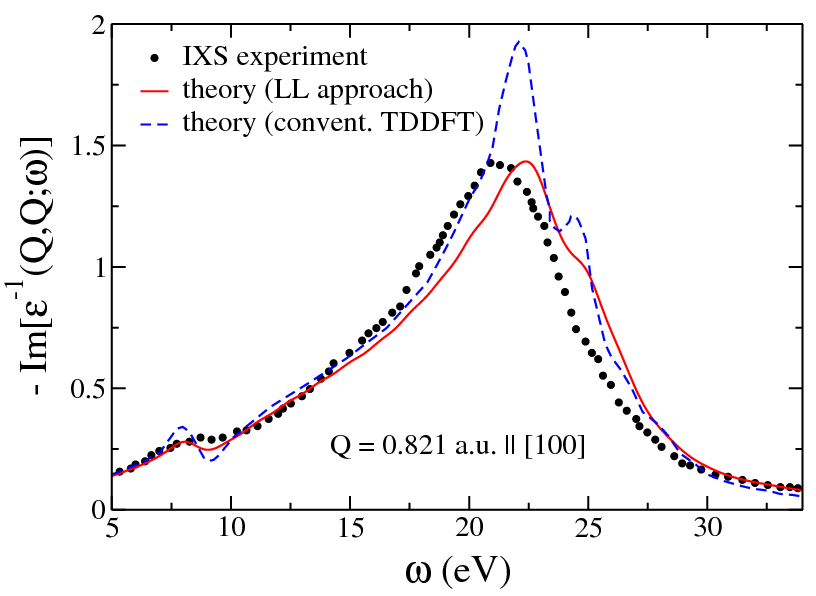}
    \label{fig:Al_theor_vs_exp_2}
  }
  \caption{Bulk Al: Comparison of the loss function calculated for two
    different values of the transferred momentum, $\mathbf{Q}$, using
    the Liouville-Lanczos (LL) approach, with
    experiment\cite{Cazzaniga:2011} and with previous
    theory.\cite{Cazzaniga:2011} (a) $Q=0.616$~a.u. along [100], (b)
    $Q=0.821$~a.u. along [100]. LL data have been obtained using 400
    Lanczos iterations plus extrapolations. A $10 \times 10 \times 10$
    ($14\times 14\times 14$) MP $\mathbf{k}$ point mesh and a
    Lorentzian broadening $\eta$ of 0.051 (0.068) Ry have been used
    for the two cases, respectively.  }
  \label{fig:Al_theor_vs_exp}
\end{figure*}

Figure~\ref{fig:Al_conv_iter} shows the convergence of the loss
function of Al, calculated at a transferred momentum
$Q=0.513$~a.u. along the [100] direction, as a function of the number
of Lanczos iterations.  Although the qualitative behavior is similar
to that observed in Si (wiggles showing up for a small number of
iterations disappear by increasing this number), the convergence
appears to be faster in the present case. As for the large-iterate
behavior of the Lanczos coefficient, we observe that, in contrast to
Si, in Al the odd and even coefficients oscillate around a same value,
which is also in this case of the order of one half the plane-wave
kinetic-energy cutoff. This is due to the vanishing of the gap, as
discussed in Ref. \onlinecite{Rocca:2008,Baroni:2012}.
Figure~\ref{fig:Al_conv_k} shows the convergence with respect to the
size of the $\mathbf{k}$ point mesh: very satisfactory convergence is
achieved with $10 \times 10 \times 10$ MP mesh and a broadening
parameter $\eta = 0.056$~Ry.
 
In Fig.~\ref{fig:Al_theor_vs_exp} we compare the loss function of Al as calculated by the
present method for two different values of the transferred momentum along the [100]
direction, with IXS experiments and with previous theoretical work. At small transferred
momentum (panel (a)) theoretical predictions agree remarkably well with each other (the
slight discrepancies being attributable to the usual small differences between the
technical details of the calculations) and with experiment.  Both theoretical spectra
display a small blueshift ($\sim 0.5$ eV) of the plasmon peak with respect to experiments.
At larger transferred momentum (panel (b)) the theoretical spectra display a feature at
$\sim 24$ eV, which is not observed experimentally. We attribute the remaining
discrepancies to the lifetime effects,\cite{Cazzaniga:2011} which have been treated in our
calculations by a constant Lorentzian broadening parameter ($\eta = 0.068$~Ry, requiring a
$14 \times 14 \times 14$ MP mesh).

\section{\label{sec:conclusions} Concluding remarks}

We believe that the Liouville-Lanczos approach introduced in this paper will open new
perspectives in the calculation of loss spectra in extended systems. Its main features are
the adoption of a representation for the charge-density response borrowed from
density-functional perturbation theory, and of a Lanczos recursion scheme for
  computing selected elements of the inverse of (very) large matrices. The combination of these two elements permits to compute the loss spectrum of a given system, for a
given transferred momentum, and for an \emph{entire wide frequency range}, with a
numerical workload of the same order as needed for a standard ground-state calculation for
a same system (the pre-factor being only a few times larger). In principle, the convergence of the computed loss spectra with respect to the length of the Lanczos chains depends on the spectral range: the lower the frequency, the faster the convergence, as it was already observed for optical spectra in finite systems.\cite{Walker:2006,Rocca:2008} In practice, however, adoption of the extrapolation techniques explained in Sec.~\ref{sec:extrapolation} substantially alleviates this dependence.  Also, the spectral range accessible to EEL/IXS spectroscopies is limited by the so-called $f$-sum rule:\cite{Mahan:1990}
\begin{equation}
  \int\limits_0^\infty \mathrm{Im} \left[
    \epsilon^{-1}(\mathbf{Q},\mathbf{Q};\omega) \right] \omega \, 
  d\omega = - \frac{\pi}{2} \, \omega_p^2 ,
  \label{eq:f_sum_rule}
\end{equation}
where $\omega_p = 4\pi e^2 n_e/m$ is the plasma frequency, $n_e$ being the average
electron density, {\it i.e.} the number of electrons (\emph{valence} electrons, in
a pseudopotential calculation) per unit volume.\cite{Timrov:Note:2013:f-sum} Of course, the spectral range that needs to be sampled by Lanczos recursion is correspondingly limited.

The Liouville-Lanczos approach introduced in this paper also lends itself to an easy
  generalization to those methods (such as hybrid functionals or the static Bethe-Salpeter
  equation -- BSE) that require the full density-matrix (rather than just charge-density)
  response, which is in fact as easily accessible to the batch representation utilized
  here.\cite{Rocca:2012} Further generalization to frequency-dependent XC kernels (or to the BSE with dynamical screening) may simply require computing the loss function at shifted
  frequencies ($\omega'=\omega+\Sigma(\omega)$), as proposed \emph{e.g.} in
  Ref. \onlinecite{Weissker:2010}, or further methodological developments. Further work is
  required to clarify this issue.

All in all we believe that the advances presented in this paper
  will allow for the simulation of complex, possibly nano-structured,
surfaces, as well as of systems where valence and shallow-core
loss spectra overlap. Examples of the former include low Miller index
surfaces or plasmonic materials, while bulk bismuth is an example of
the latter. Work is in progress on both lines.

\section*{ACKNOWLEDGMENTS}
We thank S.~de~Gironcoli, A.~Dal~Corso, and L.~Reining for valuable
discussions. Support from the ANR (Project PNANO ACCATTONE) and from
DGA are gratefully acknowledged. Computer time was granted by GENCI
(Project No. 2210). The work of I.T. and N.V. has been performed under
the auspices of the \emph{Laboratoire d'excellence en nanosciences et
  nanotechnologies Labex Nanosaclay}. N.V. thanks Marco Saitta for
discussions about TDDFPT at an early stage of the project. S.B.
gratefully acknowledges hospitality at the \emph{Laboratoire des
  Solides Irradi\'es} of the \'Ecole Polytechnique, where this paper
was written.


\begin{thebibliography}{81}
\expandafter\ifx\csname natexlab\endcsname\relax\def\natexlab#1{#1}\fi
\expandafter\ifx\csname bibnamefont\endcsname\relax
  \def\bibnamefont#1{#1}\fi
\expandafter\ifx\csname bibfnamefont\endcsname\relax
  \def\bibfnamefont#1{#1}\fi
\expandafter\ifx\csname citenamefont\endcsname\relax
  \def\citenamefont#1{#1}\fi
\expandafter\ifx\csname url\endcsname\relax
  \def\url#1{\texttt{#1}}\fi
\expandafter\ifx\csname urlprefix\endcsname\relax\def\urlprefix{URL }\fi
\providecommand{\bibinfo}[2]{#2}
\providecommand{\eprint}[2][]{\url{#2}}

\bibitem[{\citenamefont{Pines}(1956)}]{Pines:1956}
\bibinfo{author}{\bibfnamefont{D.}~\bibnamefont{Pines}}, \bibinfo{journal}{Can.
  J. Phys.} \textbf{\bibinfo{volume}{34}}, \bibinfo{pages}{1379}
  (\bibinfo{year}{1956}).

\bibitem[{\citenamefont{Nozi\`eres and Pines}(1959)}]{Nozieres:1959}
\bibinfo{author}{\bibfnamefont{P.}~\bibnamefont{Nozi\`eres}} \bibnamefont{and}
  \bibinfo{author}{\bibfnamefont{D.}~\bibnamefont{Pines}},
  \bibinfo{journal}{Phys. Rev.} \textbf{\bibinfo{volume}{113}},
  \bibinfo{pages}{1254} (\bibinfo{year}{1959}).

\bibitem[{\citenamefont{Pines}(1964)}]{Pines:1964}
\bibinfo{author}{\bibfnamefont{D.}~\bibnamefont{Pines}},
  \emph{\bibinfo{title}{Elementary excitations in solids}}
  (\bibinfo{address}{New York}, \bibinfo{year}{1964}).

\bibitem[{\citenamefont{Egerton}(1996)}]{Egerton:1996}
\bibinfo{author}{\bibfnamefont{R.~F.} \bibnamefont{Egerton}},
  \emph{\bibinfo{title}{Electron Energy-Loss Spectroscopy in the Electron
  Microscope}} (\bibinfo{publisher}{Plenum}, \bibinfo{address}{New York and
  London}, \bibinfo{year}{1996}), \bibinfo{edition}{2nd} ed.

\bibitem[{\citenamefont{Sch\"ulke}(2007)}]{Schulke:2007}
\bibinfo{author}{\bibfnamefont{W.}~\bibnamefont{Sch\"ulke}},
  \emph{\bibinfo{title}{Electron Dynamics by Inelastic {X}-Ray Scattering}}
  (\bibinfo{publisher}{Oxford University Press, Oxford}, \bibinfo{year}{2007}).

\bibitem[{\citenamefont{Atwater and Polman}(2011)}]{Atwater:2011}
\bibinfo{author}{\bibfnamefont{H.~A.} \bibnamefont{Atwater}} \bibnamefont{and}
  \bibinfo{author}{\bibfnamefont{A.}~\bibnamefont{Polman}},
  \bibinfo{journal}{Nat. Mater.} \textbf{\bibinfo{volume}{9}},
  \bibinfo{pages}{205} (\bibinfo{year}{2011}).

\bibitem[{\citenamefont{Bartal et~al.}(2011)\citenamefont{Bartal, Foord,
  Bellei, Key, Flippo, Gaillard, Offermann, Patel, Jarrott, Higginson
  et~al.}}]{Bartal:2011}
\bibinfo{author}{\bibfnamefont{T.}~\bibnamefont{Bartal}},
  \bibinfo{author}{\bibfnamefont{M.}~\bibnamefont{Foord}},
  \bibinfo{author}{\bibfnamefont{C.}~\bibnamefont{Bellei}},
  \bibinfo{author}{\bibfnamefont{M.}~\bibnamefont{Key}},
  \bibinfo{author}{\bibfnamefont{K.}~\bibnamefont{Flippo}},
  \bibinfo{author}{\bibfnamefont{S.}~\bibnamefont{Gaillard}},
  \bibinfo{author}{\bibfnamefont{D.}~\bibnamefont{Offermann}},
  \bibinfo{author}{\bibfnamefont{P.}~\bibnamefont{Patel}},
  \bibinfo{author}{\bibfnamefont{L.}~\bibnamefont{Jarrott}},
  \bibinfo{author}{\bibfnamefont{D.}~\bibnamefont{Higginson}},
  \bibnamefont{et~al.}, \bibinfo{journal}{Nature Physics}
  \textbf{\bibinfo{volume}{8}}, \bibinfo{pages}{139} (\bibinfo{year}{2011}).

\bibitem[{\citenamefont{Anker et~al.}(2008)\citenamefont{Anker, Hall, Lyandres,
  Shah, Zhao, and Duyne}}]{Anker:2008}
\bibinfo{author}{\bibfnamefont{J.~N.} \bibnamefont{Anker}},
  \bibinfo{author}{\bibfnamefont{W.~P.} \bibnamefont{Hall}},
  \bibinfo{author}{\bibfnamefont{O.}~\bibnamefont{Lyandres}},
  \bibinfo{author}{\bibfnamefont{N.~C.} \bibnamefont{Shah}},
  \bibinfo{author}{\bibfnamefont{J.}~\bibnamefont{Zhao}}, \bibnamefont{and}
  \bibinfo{author}{\bibfnamefont{R.~P.~V.} \bibnamefont{Duyne}},
  \bibinfo{journal}{Nat. Mater.} \textbf{\bibinfo{volume}{7}},
  \bibinfo{pages}{442} (\bibinfo{year}{2008}).

\bibitem[{\citenamefont{Silkin et~al.}(2005)\citenamefont{Silkin, Pitarke,
  Chulkov, and Echenique}}]{Silkin:2005}
\bibinfo{author}{\bibfnamefont{V.~M.} \bibnamefont{Silkin}},
  \bibinfo{author}{\bibfnamefont{J.~M.} \bibnamefont{Pitarke}},
  \bibinfo{author}{\bibfnamefont{E.~V.} \bibnamefont{Chulkov}},
  \bibnamefont{and} \bibinfo{author}{\bibfnamefont{P.~M.}
  \bibnamefont{Echenique}}, \bibinfo{journal}{Phys. Rev. B}
  \textbf{\bibinfo{volume}{72}}, \bibinfo{pages}{115435}
  (\bibinfo{year}{2005}).

\bibitem[{\citenamefont{Diaconescu et~al.}(2007)\citenamefont{Diaconescu, Pohl,
  Vattuone, Savio, Hofmann, Silkin, Pitarke, Chulkov, Echenique, Farias
  et~al.}}]{Diaconescu:2007}
\bibinfo{author}{\bibfnamefont{B.}~\bibnamefont{Diaconescu}},
  \bibinfo{author}{\bibfnamefont{K.}~\bibnamefont{Pohl}},
  \bibinfo{author}{\bibfnamefont{L.}~\bibnamefont{Vattuone}},
  \bibinfo{author}{\bibfnamefont{L.}~\bibnamefont{Savio}},
  \bibinfo{author}{\bibfnamefont{P.}~\bibnamefont{Hofmann}},
  \bibinfo{author}{\bibfnamefont{V.}~\bibnamefont{Silkin}},
  \bibinfo{author}{\bibfnamefont{J.}~\bibnamefont{Pitarke}},
  \bibinfo{author}{\bibfnamefont{E.}~\bibnamefont{Chulkov}},
  \bibinfo{author}{\bibfnamefont{P.}~\bibnamefont{Echenique}},
  \bibinfo{author}{\bibfnamefont{D.}~\bibnamefont{Farias}},
  \bibnamefont{et~al.}, \bibinfo{journal}{Nature}
  \textbf{\bibinfo{volume}{448}}, \bibinfo{pages}{57} (\bibinfo{year}{2007}).

\bibitem[{\citenamefont{Pohl et~al.}(2010)\citenamefont{Pohl, Diaconescu,
  Vercelli, Vattuone, Silkin, Chulkov, Echenique, and Rocca}}]{Pohl:2010}
\bibinfo{author}{\bibfnamefont{K.}~\bibnamefont{Pohl}},
  \bibinfo{author}{\bibfnamefont{B.}~\bibnamefont{Diaconescu}},
  \bibinfo{author}{\bibfnamefont{G.}~\bibnamefont{Vercelli}},
  \bibinfo{author}{\bibfnamefont{L.}~\bibnamefont{Vattuone}},
  \bibinfo{author}{\bibfnamefont{V.}~\bibnamefont{Silkin}},
  \bibinfo{author}{\bibfnamefont{E.}~\bibnamefont{Chulkov}},
  \bibinfo{author}{\bibfnamefont{P.}~\bibnamefont{Echenique}},
  \bibnamefont{and} \bibinfo{author}{\bibfnamefont{M.}~\bibnamefont{Rocca}},
  \bibinfo{journal}{European Physics Letters} \textbf{\bibinfo{volume}{90}},
  \bibinfo{pages}{57006} (\bibinfo{year}{2010}).

\bibitem[{\citenamefont{Vattuone et~al.}(2012)\citenamefont{Vattuone, Vercelli,
  Smerieri, Savio, and Rocca}}]{Vattuone:2012}
\bibinfo{author}{\bibfnamefont{L.}~\bibnamefont{Vattuone}},
  \bibinfo{author}{\bibfnamefont{G.}~\bibnamefont{Vercelli}},
  \bibinfo{author}{\bibfnamefont{M.}~\bibnamefont{Smerieri}},
  \bibinfo{author}{\bibfnamefont{L.}~\bibnamefont{Savio}}, \bibnamefont{and}
  \bibinfo{author}{\bibfnamefont{M.}~\bibnamefont{Rocca}},
  \bibinfo{journal}{Plasmonics} \textbf{\bibinfo{volume}{7}},
  \bibinfo{pages}{323} (\bibinfo{year}{2012}).

\bibitem[{\citenamefont{Yan et~al.}(2012)\citenamefont{Yan, Jacobsen, and
  Thygesen}}]{Yan:2012}
\bibinfo{author}{\bibfnamefont{J.}~\bibnamefont{Yan}},
  \bibinfo{author}{\bibfnamefont{K.}~\bibnamefont{Jacobsen}}, \bibnamefont{and}
  \bibinfo{author}{\bibfnamefont{K.}~\bibnamefont{Thygesen}},
  \bibinfo{journal}{Phys. Rev. B} \textbf{\bibinfo{volume}{86}},
  \bibinfo{pages}{241404(R)} (\bibinfo{year}{2012}).

\bibitem[{\citenamefont{Vattuone et~al.}(2013)\citenamefont{Vattuone, Smerieri,
  Langer, Tegenkamp, Pfn\"ur, Silkin, Chulkov, Echenique, and
  Rocca}}]{Vattuone:2013}
\bibinfo{author}{\bibfnamefont{L.}~\bibnamefont{Vattuone}},
  \bibinfo{author}{\bibfnamefont{M.}~\bibnamefont{Smerieri}},
  \bibinfo{author}{\bibfnamefont{T.}~\bibnamefont{Langer}},
  \bibinfo{author}{\bibfnamefont{C.}~\bibnamefont{Tegenkamp}},
  \bibinfo{author}{\bibfnamefont{H.}~\bibnamefont{Pfn\"ur}},
  \bibinfo{author}{\bibfnamefont{V.~M.} \bibnamefont{Silkin}},
  \bibinfo{author}{\bibfnamefont{E.~V.} \bibnamefont{Chulkov}},
  \bibinfo{author}{\bibfnamefont{P.~M.} \bibnamefont{Echenique}},
  \bibnamefont{and} \bibinfo{author}{\bibfnamefont{M.}~\bibnamefont{Rocca}},
  \bibinfo{journal}{Phys. Rev. Lett.} \textbf{\bibinfo{volume}{110}},
  \bibinfo{pages}{127405} (\bibinfo{year}{2013}).

\bibitem[{\citenamefont{Silkin et~al.}(2004)\citenamefont{Silkin,
  Garc\'ia-Lekue, Pitarke, Chulkov, Zaremba, and Echenique}}]{Silkin:2004}
\bibinfo{author}{\bibfnamefont{V.~M.} \bibnamefont{Silkin}},
  \bibinfo{author}{\bibfnamefont{A.}~\bibnamefont{Garc\'ia-Lekue}},
  \bibinfo{author}{\bibfnamefont{J.~M.} \bibnamefont{Pitarke}},
  \bibinfo{author}{\bibfnamefont{E.~V.} \bibnamefont{Chulkov}},
  \bibinfo{author}{\bibfnamefont{E.}~\bibnamefont{Zaremba}}, \bibnamefont{and}
  \bibinfo{author}{\bibfnamefont{P.~M.} \bibnamefont{Echenique}},
  \bibinfo{journal}{Europhys. Lett.} \textbf{\bibinfo{volume}{66}},
  \bibinfo{pages}{260} (\bibinfo{year}{2004}).

\bibitem[{\citenamefont{Pitarke et~al.}(2004)\citenamefont{Pitarke, Nazarov,
  Silkin, Chulkov, Zaremba, and Echenique}}]{Pitarke:2004}
\bibinfo{author}{\bibfnamefont{J.}~\bibnamefont{Pitarke}},
  \bibinfo{author}{\bibfnamefont{V.~U.} \bibnamefont{Nazarov}},
  \bibinfo{author}{\bibfnamefont{V.}~\bibnamefont{Silkin}},
  \bibinfo{author}{\bibfnamefont{E.}~\bibnamefont{Chulkov}},
  \bibinfo{author}{\bibfnamefont{E.}~\bibnamefont{Zaremba}}, \bibnamefont{and}
  \bibinfo{author}{\bibfnamefont{P.}~\bibnamefont{Echenique}},
  \bibinfo{journal}{Phys. Rev. B} \textbf{\bibinfo{volume}{70}},
  \bibinfo{pages}{205403} (\bibinfo{year}{2004}).

\bibitem[{\citenamefont{Pitarke et~al.}(2007)\citenamefont{Pitarke, Silkin,
  Chulkov, and Echenique}}]{Pitarke:2007}
\bibinfo{author}{\bibfnamefont{J.}~\bibnamefont{Pitarke}},
  \bibinfo{author}{\bibfnamefont{V.}~\bibnamefont{Silkin}},
  \bibinfo{author}{\bibfnamefont{E.}~\bibnamefont{Chulkov}}, \bibnamefont{and}
  \bibinfo{author}{\bibfnamefont{P.}~\bibnamefont{Echenique}},
  \bibinfo{journal}{Rep. Prog. Phys.} \textbf{\bibinfo{volume}{70}},
  \bibinfo{pages}{1} (\bibinfo{year}{2007}).

\bibitem[{\citenamefont{Esteban et~al.}(2012)\citenamefont{Esteban, Borisov,
  Nordlander, and Aizpurua}}]{Esteban:2012}
\bibinfo{author}{\bibfnamefont{R.}~\bibnamefont{Esteban}},
  \bibinfo{author}{\bibfnamefont{A.}~\bibnamefont{Borisov}},
  \bibinfo{author}{\bibfnamefont{P.}~\bibnamefont{Nordlander}},
  \bibnamefont{and} \bibinfo{author}{\bibfnamefont{J.}~\bibnamefont{Aizpurua}},
  \bibinfo{journal}{Nature Communications} \textbf{\bibinfo{volume}{3}},
  \bibinfo{pages}{825} (\bibinfo{year}{2012}).

\bibitem[{\citenamefont{Sturm}(1978)}]{Sturm:1978}
\bibinfo{author}{\bibfnamefont{K.}~\bibnamefont{Sturm}},
  \bibinfo{journal}{Phys. Rev. Lett.} \textbf{\bibinfo{volume}{40}},
  \bibinfo{pages}{1599} (\bibinfo{year}{1978}).

\bibitem[{\citenamefont{Cohen and Bergstresser}(1966)}]{Cohen:1966}
\bibinfo{author}{\bibfnamefont{M.}~\bibnamefont{Cohen}} \bibnamefont{and}
  \bibinfo{author}{\bibfnamefont{T.~K.} \bibnamefont{Bergstresser}},
  \bibinfo{journal}{Phys. Rev.} \textbf{\bibinfo{volume}{141}},
  \bibinfo{pages}{789} (\bibinfo{year}{1966}).

\bibitem[{\citenamefont{Louie et~al.}(1975)\citenamefont{Louie, Chelikowsky,
  and Cohen}}]{Louie:1975}
\bibinfo{author}{\bibfnamefont{S.~G.} \bibnamefont{Louie}},
  \bibinfo{author}{\bibfnamefont{J.}~\bibnamefont{Chelikowsky}},
  \bibnamefont{and} \bibinfo{author}{\bibfnamefont{M.~L.} \bibnamefont{Cohen}},
  \bibinfo{journal}{Phys. Rev. Lett.} \textbf{\bibinfo{volume}{34}},
  \bibinfo{pages}{155} (\bibinfo{year}{1975}).

\bibitem[{\citenamefont{Runge and Gross}(1984)}]{Runge:1984}
\bibinfo{author}{\bibfnamefont{E.}~\bibnamefont{Runge}} \bibnamefont{and}
  \bibinfo{author}{\bibfnamefont{E.}~\bibnamefont{Gross}},
  \bibinfo{journal}{Phys. Rev. Lett.} \textbf{\bibinfo{volume}{52}},
  \bibinfo{pages}{997} (\bibinfo{year}{1984}).

\bibitem[{\citenamefont{Gross et~al.}(1996)\citenamefont{Gross, Dobson, and
  Petersilka}}]{Gross:1996}
\bibinfo{author}{\bibfnamefont{E.~K.~U.} \bibnamefont{Gross}},
  \bibinfo{author}{\bibfnamefont{J.~F.} \bibnamefont{Dobson}},
  \bibnamefont{and}
  \bibinfo{author}{\bibfnamefont{M.}~\bibnamefont{Petersilka}},
  \emph{\bibinfo{title}{Density Functional Theory of Time-Dependent
  Phenomena}}, Topics in Current Chemistry
  (\bibinfo{publisher}{Springer-Verlag}, \bibinfo{address}{Berlin},
  \bibinfo{year}{1996}).

\bibitem[{\citenamefont{Caliebe et~al.}(2000)\citenamefont{Caliebe, Soininen,
  Shirley, Kao, and H\"am\"al\"ainen}}]{Caliebe:2000}
\bibinfo{author}{\bibfnamefont{W.}~\bibnamefont{Caliebe}},
  \bibinfo{author}{\bibfnamefont{J.}~\bibnamefont{Soininen}},
  \bibinfo{author}{\bibfnamefont{E.}~\bibnamefont{Shirley}},
  \bibinfo{author}{\bibfnamefont{C.-C.} \bibnamefont{Kao}}, \bibnamefont{and}
  \bibinfo{author}{\bibfnamefont{K.}~\bibnamefont{H\"am\"al\"ainen}},
  \bibinfo{journal}{Phys. Rev. Lett.} \textbf{\bibinfo{volume}{84}},
  \bibinfo{pages}{3907} (\bibinfo{year}{2000}).

\bibitem[{\citenamefont{Olevano and Reining}(2001)}]{Olevano:2001}
\bibinfo{author}{\bibfnamefont{V.}~\bibnamefont{Olevano}} \bibnamefont{and}
  \bibinfo{author}{\bibfnamefont{L.}~\bibnamefont{Reining}},
  \bibinfo{journal}{Phys. Rev. Lett.} \textbf{\bibinfo{volume}{86}},
  \bibinfo{pages}{5962} (\bibinfo{year}{2001}).

\bibitem[{\citenamefont{Takada and Yasuhara}(2002)}]{Takada:2002}
\bibinfo{author}{\bibfnamefont{Y.}~\bibnamefont{Takada}} \bibnamefont{and}
  \bibinfo{author}{\bibfnamefont{H.}~\bibnamefont{Yasuhara}},
  \bibinfo{journal}{Phys. Rev. Lett.} \textbf{\bibinfo{volume}{89}},
  \bibinfo{pages}{216402} (\bibinfo{year}{2002}).

\bibitem[{\citenamefont{Arnaud et~al.}(2005)\citenamefont{Arnaud, Leb\`egue,
  and Alouani}}]{Arnaud:2005}
\bibinfo{author}{\bibfnamefont{B.}~\bibnamefont{Arnaud}},
  \bibinfo{author}{\bibfnamefont{S.}~\bibnamefont{Leb\`egue}},
  \bibnamefont{and} \bibinfo{author}{\bibfnamefont{M.}~\bibnamefont{Alouani}},
  \bibinfo{journal}{Phys. Rev. B} \textbf{\bibinfo{volume}{71}},
  \bibinfo{pages}{035308} (\bibinfo{year}{2005}).

\bibitem[{\citenamefont{Daling et~al.}(1992)\citenamefont{Daling, van
  Haeringen, and Farid}}]{Daling:1992}
\bibinfo{author}{\bibfnamefont{R.}~\bibnamefont{Daling}},
  \bibinfo{author}{\bibfnamefont{W.}~\bibnamefont{van Haeringen}},
  \bibnamefont{and} \bibinfo{author}{\bibfnamefont{B.}~\bibnamefont{Farid}},
  \bibinfo{journal}{Phys. Rev. B} \textbf{\bibinfo{volume}{45}},
  \bibinfo{pages}{8970} (\bibinfo{year}{1992}).

\bibitem[{\citenamefont{Engel and Farid}(1992)}]{Engel:1992}
\bibinfo{author}{\bibfnamefont{G.}~\bibnamefont{Engel}} \bibnamefont{and}
  \bibinfo{author}{\bibfnamefont{B.}~\bibnamefont{Farid}},
  \bibinfo{journal}{Phys. Rev. B} \textbf{\bibinfo{volume}{46}},
  \bibinfo{pages}{15812} (\bibinfo{year}{1992}).

\bibitem[{\citenamefont{Sturm et~al.}(1992)\citenamefont{Sturm, Sch\"ulke, and
  Schmitz}}]{Sturm:1992}
\bibinfo{author}{\bibfnamefont{K.}~\bibnamefont{Sturm}},
  \bibinfo{author}{\bibfnamefont{W.}~\bibnamefont{Sch\"ulke}},
  \bibnamefont{and} \bibinfo{author}{\bibfnamefont{J.}~\bibnamefont{Schmitz}},
  \bibinfo{journal}{Phys. Rev. Lett.} \textbf{\bibinfo{volume}{68}},
  \bibinfo{pages}{228} (\bibinfo{year}{1992}).

\bibitem[{\citenamefont{Quong and Eguiluz}(1993)}]{Quong:1993}
\bibinfo{author}{\bibfnamefont{A.~A.} \bibnamefont{Quong}} \bibnamefont{and}
  \bibinfo{author}{\bibfnamefont{A.}~\bibnamefont{Eguiluz}},
  \bibinfo{journal}{Phys. Rev. Lett.} \textbf{\bibinfo{volume}{70}},
  \bibinfo{pages}{3955} (\bibinfo{year}{1993}).

\bibitem[{\citenamefont{Fleszar et~al.}(1995)\citenamefont{Fleszar, Quong, and
  Eguiluz}}]{Fleszar:1995}
\bibinfo{author}{\bibfnamefont{A.}~\bibnamefont{Fleszar}},
  \bibinfo{author}{\bibfnamefont{A.}~\bibnamefont{Quong}}, \bibnamefont{and}
  \bibinfo{author}{\bibfnamefont{A.}~\bibnamefont{Eguiluz}},
  \bibinfo{journal}{Phys. Rev. Lett.} \textbf{\bibinfo{volume}{74}},
  \bibinfo{pages}{590} (\bibinfo{year}{1995}).

\bibitem[{\citenamefont{Ehrnsperger and Bross}(1997)}]{Ehrnsperger:1997}
\bibinfo{author}{\bibfnamefont{M.}~\bibnamefont{Ehrnsperger}} \bibnamefont{and}
  \bibinfo{author}{\bibfnamefont{H.}~\bibnamefont{Bross}}, \bibinfo{journal}{J.
  Phys.: Condens. Matter} \textbf{\bibinfo{volume}{9}}, \bibinfo{pages}{1225}
  (\bibinfo{year}{1997}).

\bibitem[{\citenamefont{Waidmann et~al.}(2000)\citenamefont{Waidmann, Knupfer,
  Arnold, Fink, Fleszar, and Hanke}}]{Waidmann:2000}
\bibinfo{author}{\bibfnamefont{S.}~\bibnamefont{Waidmann}},
  \bibinfo{author}{\bibfnamefont{M.}~\bibnamefont{Knupfer}},
  \bibinfo{author}{\bibfnamefont{B.}~\bibnamefont{Arnold}},
  \bibinfo{author}{\bibfnamefont{J.}~\bibnamefont{Fink}},
  \bibinfo{author}{\bibfnamefont{A.}~\bibnamefont{Fleszar}}, \bibnamefont{and}
  \bibinfo{author}{\bibfnamefont{W.}~\bibnamefont{Hanke}},
  \bibinfo{journal}{Phys. Rev. B} \textbf{\bibinfo{volume}{61}},
  \bibinfo{pages}{10149} (\bibinfo{year}{2000}).

\bibitem[{\citenamefont{Vast et~al.}(2002)\citenamefont{Vast, Reining, Olevano,
  Schattschneider, and Jouffrey}}]{Vast:2002}
\bibinfo{author}{\bibfnamefont{N.}~\bibnamefont{Vast}},
  \bibinfo{author}{\bibfnamefont{L.}~\bibnamefont{Reining}},
  \bibinfo{author}{\bibfnamefont{V.}~\bibnamefont{Olevano}},
  \bibinfo{author}{\bibfnamefont{P.}~\bibnamefont{Schattschneider}},
  \bibnamefont{and} \bibinfo{author}{\bibfnamefont{B.}~\bibnamefont{Jouffrey}},
  \bibinfo{journal}{Phys. Rev. Lett.} \textbf{\bibinfo{volume}{88}},
  \bibinfo{pages}{037601} (\bibinfo{year}{2002}).

\bibitem[{\citenamefont{Marinopoulos et~al.}(2002)\citenamefont{Marinopoulos,
  Reining, Olevano, Rubio, Pichler, Liu, Knupfer, and
  Fink}}]{Marinopoulos:2002}
\bibinfo{author}{\bibfnamefont{A.}~\bibnamefont{Marinopoulos}},
  \bibinfo{author}{\bibfnamefont{L.}~\bibnamefont{Reining}},
  \bibinfo{author}{\bibfnamefont{V.}~\bibnamefont{Olevano}},
  \bibinfo{author}{\bibfnamefont{A.}~\bibnamefont{Rubio}},
  \bibinfo{author}{\bibfnamefont{T.}~\bibnamefont{Pichler}},
  \bibinfo{author}{\bibfnamefont{X.}~\bibnamefont{Liu}},
  \bibinfo{author}{\bibfnamefont{M.}~\bibnamefont{Knupfer}}, \bibnamefont{and}
  \bibinfo{author}{\bibfnamefont{J.}~\bibnamefont{Fink}},
  \bibinfo{journal}{Phys. Rev. Lett.} \textbf{\bibinfo{volume}{89}},
  \bibinfo{pages}{076402} (\bibinfo{year}{2002}).

\bibitem[{\citenamefont{Marinopoulos et~al.}(2003)\citenamefont{Marinopoulos,
  Reining, Rubio, and Vast}}]{Marinopoulos:2003}
\bibinfo{author}{\bibfnamefont{A.}~\bibnamefont{Marinopoulos}},
  \bibinfo{author}{\bibfnamefont{L.}~\bibnamefont{Reining}},
  \bibinfo{author}{\bibfnamefont{A.}~\bibnamefont{Rubio}}, \bibnamefont{and}
  \bibinfo{author}{\bibfnamefont{N.}~\bibnamefont{Vast}},
  \bibinfo{journal}{Phys. Rev. Lett.} \textbf{\bibinfo{volume}{91}},
  \bibinfo{pages}{046402} (\bibinfo{year}{2003}).

\bibitem[{\citenamefont{Sch\"one et~al.}(2003)\citenamefont{Sch\"one, Su, and
  Ekardt}}]{Schone:2003}
\bibinfo{author}{\bibfnamefont{W.-D.} \bibnamefont{Sch\"one}},
  \bibinfo{author}{\bibfnamefont{D.~S.} \bibnamefont{Su}}, \bibnamefont{and}
  \bibinfo{author}{\bibfnamefont{W.}~\bibnamefont{Ekardt}},
  \bibinfo{journal}{Phys. Rev. B} \textbf{\bibinfo{volume}{68}},
  \bibinfo{pages}{115102} (\bibinfo{year}{2003}).

\bibitem[{\citenamefont{Dash et~al.}(2004)\citenamefont{Dash, Vast, Baranek,
  Cheynet, and Reining}}]{Dash:2004}
\bibinfo{author}{\bibfnamefont{L.}~\bibnamefont{Dash}},
  \bibinfo{author}{\bibfnamefont{N.}~\bibnamefont{Vast}},
  \bibinfo{author}{\bibfnamefont{P.}~\bibnamefont{Baranek}},
  \bibinfo{author}{\bibfnamefont{M.-C.} \bibnamefont{Cheynet}},
  \bibnamefont{and} \bibinfo{author}{\bibfnamefont{L.}~\bibnamefont{Reining}},
  \bibinfo{journal}{Physical Review B} \textbf{\bibinfo{volume}{70}},
  \bibinfo{pages}{245116} (\bibinfo{year}{2004}).

\bibitem[{\citenamefont{Gurtubay et~al.}(2004)\citenamefont{Gurtubay, Ku,
  Pitarke, Eguiluz, Larson, Tischler, and Zschack}}]{Gurtubay:2004}
\bibinfo{author}{\bibfnamefont{I.~G.} \bibnamefont{Gurtubay}},
  \bibinfo{author}{\bibfnamefont{W.}~\bibnamefont{Ku}},
  \bibinfo{author}{\bibfnamefont{J.~M.} \bibnamefont{Pitarke}},
  \bibinfo{author}{\bibfnamefont{A.~G.} \bibnamefont{Eguiluz}},
  \bibinfo{author}{\bibfnamefont{B.~C.} \bibnamefont{Larson}},
  \bibinfo{author}{\bibfnamefont{J.}~\bibnamefont{Tischler}}, \bibnamefont{and}
  \bibinfo{author}{\bibfnamefont{P.}~\bibnamefont{Zschack}},
  \bibinfo{journal}{Phys. Rev. B} \textbf{\bibinfo{volume}{70}},
  \bibinfo{pages}{201201(R)} (\bibinfo{year}{2004}).

\bibitem[{\citenamefont{Gurtubay et~al.}(2005)\citenamefont{Gurtubay, Pitarke,
  Ku, Eguiluz, Larson, Tischler, Zschack, and Finkelstein}}]{Gurtubay:2005}
\bibinfo{author}{\bibfnamefont{I.~G.} \bibnamefont{Gurtubay}},
  \bibinfo{author}{\bibfnamefont{J.~M.} \bibnamefont{Pitarke}},
  \bibinfo{author}{\bibfnamefont{W.}~\bibnamefont{Ku}},
  \bibinfo{author}{\bibfnamefont{A.~G.} \bibnamefont{Eguiluz}},
  \bibinfo{author}{\bibfnamefont{B.~C.} \bibnamefont{Larson}},
  \bibinfo{author}{\bibfnamefont{J.}~\bibnamefont{Tischler}},
  \bibinfo{author}{\bibfnamefont{P.}~\bibnamefont{Zschack}}, \bibnamefont{and}
  \bibinfo{author}{\bibfnamefont{K.~D.} \bibnamefont{Finkelstein}},
  \bibinfo{journal}{Phys. Rev. B} \textbf{\bibinfo{volume}{72}},
  \bibinfo{pages}{125117} (\bibinfo{year}{2005}).

\bibitem[{\citenamefont{Weissker et~al.}(2006)\citenamefont{Weissker, Serrano,
  Huotari, Bruneval, Sottile, Monaco, Krisch, Olevano, and
  Reining}}]{Weissker:2006}
\bibinfo{author}{\bibfnamefont{H.-C.} \bibnamefont{Weissker}},
  \bibinfo{author}{\bibfnamefont{J.}~\bibnamefont{Serrano}},
  \bibinfo{author}{\bibfnamefont{S.}~\bibnamefont{Huotari}},
  \bibinfo{author}{\bibfnamefont{F.}~\bibnamefont{Bruneval}},
  \bibinfo{author}{\bibfnamefont{F.}~\bibnamefont{Sottile}},
  \bibinfo{author}{\bibfnamefont{G.}~\bibnamefont{Monaco}},
  \bibinfo{author}{\bibfnamefont{M.}~\bibnamefont{Krisch}},
  \bibinfo{author}{\bibfnamefont{V.}~\bibnamefont{Olevano}}, \bibnamefont{and}
  \bibinfo{author}{\bibfnamefont{L.}~\bibnamefont{Reining}},
  \bibinfo{journal}{Phys. Rev. Lett.} \textbf{\bibinfo{volume}{97}},
  \bibinfo{pages}{237602} (\bibinfo{year}{2006}).

\bibitem[{\citenamefont{Kramberger et~al.}(2008)\citenamefont{Kramberger,
  Hambach, Giorgetti, R\"ummeli, Knupfer, Fink, b\"uchner, Reining, Einarsson,
  Maruyama et~al.}}]{Kramberger:2008}
\bibinfo{author}{\bibfnamefont{C.}~\bibnamefont{Kramberger}},
  \bibinfo{author}{\bibfnamefont{R.}~\bibnamefont{Hambach}},
  \bibinfo{author}{\bibfnamefont{C.}~\bibnamefont{Giorgetti}},
  \bibinfo{author}{\bibfnamefont{M.}~\bibnamefont{R\"ummeli}},
  \bibinfo{author}{\bibfnamefont{M.}~\bibnamefont{Knupfer}},
  \bibinfo{author}{\bibfnamefont{J.}~\bibnamefont{Fink}},
  \bibinfo{author}{\bibfnamefont{B.}~\bibnamefont{b\"uchner}},
  \bibinfo{author}{\bibfnamefont{L.}~\bibnamefont{Reining}},
  \bibinfo{author}{\bibfnamefont{E.}~\bibnamefont{Einarsson}},
  \bibinfo{author}{\bibfnamefont{S.}~\bibnamefont{Maruyama}},
  \bibnamefont{et~al.}, \bibinfo{journal}{Phys. Rev. Lett.}
  \textbf{\bibinfo{volume}{100}}, \bibinfo{pages}{196803}
  (\bibinfo{year}{2008}).

\bibitem[{\citenamefont{Alkauskas et~al.}(2010)\citenamefont{Alkauskas,
  Schneider, Sagmeister, Ambrosch-Draxl, and H\'ebert}}]{Alkauskas:2010}
\bibinfo{author}{\bibfnamefont{A.}~\bibnamefont{Alkauskas}},
  \bibinfo{author}{\bibfnamefont{S.}~\bibnamefont{Schneider}},
  \bibinfo{author}{\bibfnamefont{S.}~\bibnamefont{Sagmeister}},
  \bibinfo{author}{\bibfnamefont{C.}~\bibnamefont{Ambrosch-Draxl}},
  \bibnamefont{and} \bibinfo{author}{\bibfnamefont{C.}~\bibnamefont{H\'ebert}},
  \bibinfo{journal}{Ultramicroscopy} \textbf{\bibinfo{volume}{110}},
  \bibinfo{pages}{1081} (\bibinfo{year}{2010}).

\bibitem[{\citenamefont{Huotari et~al.}(2009)\citenamefont{Huotari, Sternemann,
  Troparevsky, Eguiluz, Volmer, Sternemann, M\"uller, Monaco, and
  Sch\"ulke}}]{Huotari:2009}
\bibinfo{author}{\bibfnamefont{S.}~\bibnamefont{Huotari}},
  \bibinfo{author}{\bibfnamefont{C.}~\bibnamefont{Sternemann}},
  \bibinfo{author}{\bibfnamefont{M.}~\bibnamefont{Troparevsky}},
  \bibinfo{author}{\bibfnamefont{A.}~\bibnamefont{Eguiluz}},
  \bibinfo{author}{\bibfnamefont{M.}~\bibnamefont{Volmer}},
  \bibinfo{author}{\bibfnamefont{H.}~\bibnamefont{Sternemann}},
  \bibinfo{author}{\bibfnamefont{H.}~\bibnamefont{M\"uller}},
  \bibinfo{author}{\bibfnamefont{G.}~\bibnamefont{Monaco}}, \bibnamefont{and}
  \bibinfo{author}{\bibfnamefont{W.}~\bibnamefont{Sch\"ulke}},
  \bibinfo{journal}{Phys. Rev. B} \textbf{\bibinfo{volume}{80}},
  \bibinfo{pages}{155107} (\bibinfo{year}{2009}).

\bibitem[{\citenamefont{Weissker et~al.}(2010)\citenamefont{Weissker, Serrano,
  Huotari, Luppi, Cazzaniga, Bruneval, Sottile, Monaco, Olevano, and
  Reining}}]{Weissker:2010}
\bibinfo{author}{\bibfnamefont{H.-C.} \bibnamefont{Weissker}},
  \bibinfo{author}{\bibfnamefont{J.}~\bibnamefont{Serrano}},
  \bibinfo{author}{\bibfnamefont{S.}~\bibnamefont{Huotari}},
  \bibinfo{author}{\bibfnamefont{E.}~\bibnamefont{Luppi}},
  \bibinfo{author}{\bibfnamefont{M.}~\bibnamefont{Cazzaniga}},
  \bibinfo{author}{\bibfnamefont{F.}~\bibnamefont{Bruneval}},
  \bibinfo{author}{\bibfnamefont{F.}~\bibnamefont{Sottile}},
  \bibinfo{author}{\bibfnamefont{G.}~\bibnamefont{Monaco}},
  \bibinfo{author}{\bibfnamefont{V.}~\bibnamefont{Olevano}}, \bibnamefont{and}
  \bibinfo{author}{\bibfnamefont{L.}~\bibnamefont{Reining}},
  \bibinfo{journal}{Phys. Rev. B} \textbf{\bibinfo{volume}{81}},
  \bibinfo{pages}{085104} (\bibinfo{year}{2010}).

\bibitem[{\citenamefont{Cazzaniga et~al.}(2011)\citenamefont{Cazzaniga,
  Weissker, Huotari, Pylkk\"anen, Salvestrini, Monaco, Onida, and
  Reining}}]{Cazzaniga:2011}
\bibinfo{author}{\bibfnamefont{M.}~\bibnamefont{Cazzaniga}},
  \bibinfo{author}{\bibfnamefont{H.-C.} \bibnamefont{Weissker}},
  \bibinfo{author}{\bibfnamefont{S.}~\bibnamefont{Huotari}},
  \bibinfo{author}{\bibfnamefont{T.}~\bibnamefont{Pylkk\"anen}},
  \bibinfo{author}{\bibfnamefont{P.}~\bibnamefont{Salvestrini}},
  \bibinfo{author}{\bibfnamefont{G.}~\bibnamefont{Monaco}},
  \bibinfo{author}{\bibfnamefont{G.}~\bibnamefont{Onida}}, \bibnamefont{and}
  \bibinfo{author}{\bibfnamefont{L.}~\bibnamefont{Reining}},
  \bibinfo{journal}{Phys. Rev. B} \textbf{\bibinfo{volume}{84}},
  \bibinfo{pages}{075109} (\bibinfo{year}{2011}).

\bibitem[{\citenamefont{Yan et~al.}(2011)\citenamefont{Yan, Thygesen, and
  Jacobsen}}]{Yan:2011}
\bibinfo{author}{\bibfnamefont{J.}~\bibnamefont{Yan}},
  \bibinfo{author}{\bibfnamefont{K.~S.} \bibnamefont{Thygesen}},
  \bibnamefont{and} \bibinfo{author}{\bibfnamefont{K.~W.}
  \bibnamefont{Jacobsen}}, \bibinfo{journal}{Phys. Rev. Lett.}
  \textbf{\bibinfo{volume}{106}}, \bibinfo{pages}{146803}
  (\bibinfo{year}{2011}).

\bibitem[{\citenamefont{Onida et~al.}(2002)\citenamefont{Onida, Reining, and
  Rubio}}]{Onida:2002}
\bibinfo{author}{\bibfnamefont{G.}~\bibnamefont{Onida}},
  \bibinfo{author}{\bibfnamefont{L.}~\bibnamefont{Reining}}, \bibnamefont{and}
  \bibinfo{author}{\bibfnamefont{A.}~\bibnamefont{Rubio}},
  \bibinfo{journal}{Rev. Mod. Phys.} \textbf{\bibinfo{volume}{74}},
  \bibinfo{pages}{601} (\bibinfo{year}{2002}).

\bibitem[{\citenamefont{Walker et~al.}(2006)\citenamefont{Walker, Saitta,
  Gebauer, and Baroni}}]{Walker:2006}
\bibinfo{author}{\bibfnamefont{B.}~\bibnamefont{Walker}},
  \bibinfo{author}{\bibfnamefont{A.~M.} \bibnamefont{Saitta}},
  \bibinfo{author}{\bibfnamefont{R.}~\bibnamefont{Gebauer}}, \bibnamefont{and}
  \bibinfo{author}{\bibfnamefont{S.}~\bibnamefont{Baroni}},
  \bibinfo{journal}{Phys. Rev. Lett.} \textbf{\bibinfo{volume}{96}},
  \bibinfo{pages}{113001} (\bibinfo{year}{2006}).

\bibitem[{\citenamefont{Rocca et~al.}(2008)\citenamefont{Rocca, Gebauer, Saas,
  and Baroni}}]{Rocca:2008}
\bibinfo{author}{\bibfnamefont{D.}~\bibnamefont{Rocca}},
  \bibinfo{author}{\bibfnamefont{R.}~\bibnamefont{Gebauer}},
  \bibinfo{author}{\bibfnamefont{Y.}~\bibnamefont{Saas}}, \bibnamefont{and}
  \bibinfo{author}{\bibfnamefont{S.}~\bibnamefont{Baroni}},
  \bibinfo{journal}{J. Chem. Phys.} \textbf{\bibinfo{volume}{128}},
  \bibinfo{pages}{154105} (\bibinfo{year}{2008}).

\bibitem[{\citenamefont{Malcioglu et~al.}(2011)\citenamefont{Malcioglu,
  Gebauer, Rocca, and Baroni}}]{Malcioglu:2011}
\bibinfo{author}{\bibfnamefont{O.~B.} \bibnamefont{Malcioglu}},
  \bibinfo{author}{\bibfnamefont{R.}~\bibnamefont{Gebauer}},
  \bibinfo{author}{\bibfnamefont{D.}~\bibnamefont{Rocca}}, \bibnamefont{and}
  \bibinfo{author}{\bibfnamefont{S.}~\bibnamefont{Baroni}},
  \bibinfo{journal}{Comput. Phys. Commun.} \textbf{\bibinfo{volume}{182}},
  \bibinfo{pages}{1744} (\bibinfo{year}{2011}).

\bibitem[{\citenamefont{Baroni and Gebauer}(2012)}]{Baroni:2012}
\bibinfo{author}{\bibfnamefont{S.}~\bibnamefont{Baroni}} \bibnamefont{and}
  \bibinfo{author}{\bibfnamefont{R.}~\bibnamefont{Gebauer}},
  \emph{\bibinfo{title}{Fundamentals of Time-Dependent Density Functional
  Theory}} (\bibinfo{publisher}{Springer}, \bibinfo{address}{Berlin},
  \bibinfo{year}{2012}).

\bibitem[{\citenamefont{Soininen et~al.}(2005)\citenamefont{Soininen,
  Ankudinov, and Rehr}}]{Soininen:2005}
\bibinfo{author}{\bibfnamefont{J.~A.} \bibnamefont{Soininen}},
  \bibinfo{author}{\bibfnamefont{A.~L.} \bibnamefont{Ankudinov}},
  \bibnamefont{and} \bibinfo{author}{\bibfnamefont{J.~J.} \bibnamefont{Rehr}},
  \bibinfo{journal}{Phys. Rev. B} \textbf{\bibinfo{volume}{72}},
  \bibinfo{pages}{045136} (\bibinfo{year}{2005}).

\bibitem[{\citenamefont{Joly}(2001)}]{Joly:2001}
\bibinfo{author}{\bibfnamefont{Y.}~\bibnamefont{Joly}}, \bibinfo{journal}{Phys.
  Rev. B} \textbf{\bibinfo{volume}{63}}, \bibinfo{pages}{125120}
  (\bibinfo{year}{2001}).

\bibitem[{\citenamefont{Cabaret et~al.}(2013)\citenamefont{Cabaret, Emery,
  Bellin, Herold, Lagrange, Wilhelm, Rogalev, and Loupias}}]{Cabaret:2013}
\bibinfo{author}{\bibfnamefont{D.}~\bibnamefont{Cabaret}},
  \bibinfo{author}{\bibfnamefont{N.}~\bibnamefont{Emery}},
  \bibinfo{author}{\bibfnamefont{C.}~\bibnamefont{Bellin}},
  \bibinfo{author}{\bibfnamefont{C.}~\bibnamefont{Herold}},
  \bibinfo{author}{\bibfnamefont{P.}~\bibnamefont{Lagrange}},
  \bibinfo{author}{\bibfnamefont{F.}~\bibnamefont{Wilhelm}},
  \bibinfo{author}{\bibfnamefont{A.}~\bibnamefont{Rogalev}}, \bibnamefont{and}
  \bibinfo{author}{\bibfnamefont{G.}~\bibnamefont{Loupias}},
  \bibinfo{journal}{Phys. Rev. B} \textbf{\bibinfo{volume}{87}},
  \bibinfo{pages}{075108} (\bibinfo{year}{2013}).

\bibitem[{\citenamefont{Baroni et~al.}(2001)\citenamefont{Baroni, de~Gironcoli,
  Dal~Corso, and Giannozzi}}]{Baroni:2001}
\bibinfo{author}{\bibfnamefont{S.}~\bibnamefont{Baroni}},
  \bibinfo{author}{\bibfnamefont{S.}~\bibnamefont{de~Gironcoli}},
  \bibinfo{author}{\bibfnamefont{A.}~\bibnamefont{Dal~Corso}},
  \bibnamefont{and}
  \bibinfo{author}{\bibfnamefont{P.}~\bibnamefont{Giannozzi}},
  \bibinfo{journal}{Rev. Mod. Phys.} \textbf{\bibinfo{volume}{73}},
  \bibinfo{pages}{515} (\bibinfo{year}{2001}).

\bibitem[{\citenamefont{Hove}(1954)}]{VanHove:1954}
\bibinfo{author}{\bibfnamefont{L.~V.} \bibnamefont{Hove}},
  \bibinfo{journal}{Phys. Rev.} \textbf{\bibinfo{volume}{95}},
  \bibinfo{pages}{249} (\bibinfo{year}{1954}).

\bibitem[{\citenamefont{Mao et~al.}(2001)\citenamefont{Mao, Kao, and
  Hemley}}]{Mao:2001}
\bibinfo{author}{\bibfnamefont{H.-K.} \bibnamefont{Mao}},
  \bibinfo{author}{\bibfnamefont{C.}~\bibnamefont{Kao}}, \bibnamefont{and}
  \bibinfo{author}{\bibfnamefont{R.}~\bibnamefont{Hemley}},
  \bibinfo{journal}{J. Phys.: Condens. Matter} \textbf{\bibinfo{volume}{13}},
  \bibinfo{pages}{7847} (\bibinfo{year}{2001}).

\bibitem[{\citenamefont{Loa et~al.}(2011)\citenamefont{Loa, Syassen, Monaco,
  Vank\`o, Krish, and Hanfland}}]{Loa:2011}
\bibinfo{author}{\bibfnamefont{I.}~\bibnamefont{Loa}},
  \bibinfo{author}{\bibfnamefont{K.}~\bibnamefont{Syassen}},
  \bibinfo{author}{\bibfnamefont{G.}~\bibnamefont{Monaco}},
  \bibinfo{author}{\bibfnamefont{G.}~\bibnamefont{Vank\`o}},
  \bibinfo{author}{\bibfnamefont{M.}~\bibnamefont{Krish}}, \bibnamefont{and}
  \bibinfo{author}{\bibfnamefont{M.}~\bibnamefont{Hanfland}},
  \bibinfo{journal}{Phys. Rev. Lett.} \textbf{\bibinfo{volume}{107}},
  \bibinfo{pages}{086402} (\bibinfo{year}{2011}).

\bibitem[{\citenamefont{Pines and Nozi\`eres}(1966)}]{Pines:1966}
\bibinfo{author}{\bibfnamefont{D.}~\bibnamefont{Pines}} \bibnamefont{and}
  \bibinfo{author}{\bibfnamefont{P.}~\bibnamefont{Nozi\`eres}},
  \emph{\bibinfo{title}{The Theory of Quantum Liquids}},
  vol.~\bibinfo{volume}{1} (\bibinfo{publisher}{Benjamin},
  \bibinfo{address}{New York}, \bibinfo{year}{1966}).

\bibitem[{\citenamefont{Martin}(2004)}]{Martin:2004}
\bibinfo{editor}{\bibfnamefont{R.}~\bibnamefont{Martin}}, ed.,
  \emph{\bibinfo{title}{Electronic Structure: Basic Theory and Practical
  Methods}} (\bibinfo{publisher}{Cambridge University Press},
  \bibinfo{address}{Cambridge}, \bibinfo{year}{2004}).

\bibitem[{\citenamefont{Car et~al.}(1981)\citenamefont{Car, Tosatti, Baroni,
  and Leelaprute}}]{Car:1981}
\bibinfo{author}{\bibfnamefont{R.}~\bibnamefont{Car}},
  \bibinfo{author}{\bibfnamefont{E.}~\bibnamefont{Tosatti}},
  \bibinfo{author}{\bibfnamefont{S.}~\bibnamefont{Baroni}}, \bibnamefont{and}
  \bibinfo{author}{\bibfnamefont{S.}~\bibnamefont{Leelaprute}},
  \bibinfo{journal}{Phys. Rev. B} \textbf{\bibinfo{volume}{24}},
  \bibinfo{pages}{985} (\bibinfo{year}{1981}).

\bibitem[{\citenamefont{Gross and Kohn}(1985)}]{Gross:1985}
\bibinfo{author}{\bibfnamefont{E.~K.~U.} \bibnamefont{Gross}} \bibnamefont{and}
  \bibinfo{author}{\bibfnamefont{W.}~\bibnamefont{Kohn}},
  \bibinfo{journal}{Phys. Rev. Lett.} \textbf{\bibinfo{volume}{55}},
  \bibinfo{pages}{2850} (\bibinfo{year}{1985}).

\bibitem[{Tim({\natexlab{a}})}]{Timrov:Note:2013:2NcxNv}
\bibinfo{note}{According to Eq.~\eqref{eq:charge-dens-response_2}, the response
  density matrix is uniquely identified by the two sets of $N_v$ response
  orbitals $\{\varphi'_v(\mathbf{r},\pm\omega)\}$, each one of which is a
  linear combination of $N_c$ virtual orbitals.}

\bibitem[{\citenamefont{Bullet et~al.}(1980)\citenamefont{Bullet, Haydock,
  Heine, and Kelly}}]{Bullet:1980}
\bibinfo{author}{\bibfnamefont{D.~W.} \bibnamefont{Bullet}},
  \bibinfo{author}{\bibfnamefont{R.}~\bibnamefont{Haydock}},
  \bibinfo{author}{\bibfnamefont{V.}~\bibnamefont{Heine}}, \bibnamefont{and}
  \bibinfo{author}{\bibfnamefont{M.}~\bibnamefont{Kelly}},
  \emph{\bibinfo{title}{Solid State Physics}}, vol.~\bibinfo{volume}{35}
  (\bibinfo{publisher}{Academic}, \bibinfo{address}{New York},
  \bibinfo{year}{1980}).

\bibitem[{\citenamefont{Saad}(2003)}]{Saad:2003}
\bibinfo{author}{\bibfnamefont{Y.}~\bibnamefont{Saad}},
  \emph{\bibinfo{title}{Iterative Methods for Sparse Linear Systems}}
  (\bibinfo{publisher}{SIAM}, \bibinfo{address}{Philadelphia},
  \bibinfo{year}{2003}), \bibinfo{edition}{2nd} ed.

\bibitem[{\citenamefont{Ankudinov et~al.}(2002)\citenamefont{Ankudinov,
  Bouldin, Rehr, Sims, and Hung}}]{Ankudinov:2002}
\bibinfo{author}{\bibfnamefont{A.~L.} \bibnamefont{Ankudinov}},
  \bibinfo{author}{\bibfnamefont{C.~E.} \bibnamefont{Bouldin}},
  \bibinfo{author}{\bibfnamefont{J.~J.} \bibnamefont{Rehr}},
  \bibinfo{author}{\bibfnamefont{J.}~\bibnamefont{Sims}}, \bibnamefont{and}
  \bibinfo{author}{\bibfnamefont{H.}~\bibnamefont{Hung}},
  \bibinfo{journal}{Phys. Rev. B} \textbf{\bibinfo{volume}{65}},
  \bibinfo{pages}{104107} (\bibinfo{year}{2002}).

\bibitem[{\citenamefont{Gr\"uning et~al.}(2011)\citenamefont{Gr\"uning, Marini,
  and Gonze}}]{Gruning:2011}
\bibinfo{author}{\bibfnamefont{M.}~\bibnamefont{Gr\"uning}},
  \bibinfo{author}{\bibfnamefont{A.}~\bibnamefont{Marini}}, \bibnamefont{and}
  \bibinfo{author}{\bibfnamefont{X.}~\bibnamefont{Gonze}},
  \bibinfo{journal}{Comp. Mater. Sci.} \textbf{\bibinfo{volume}{50}},
  \bibinfo{pages}{2148} (\bibinfo{year}{2011}).

\bibitem[{\citenamefont{Timrov}(2013)}]{Timrov:2013}
\bibinfo{author}{\bibfnamefont{I.}~\bibnamefont{Timrov}}, Ph.D. thesis,
  \bibinfo{school}{\'Ecole Polytechnique}, \bibinfo{address}{France}
  (\bibinfo{year}{2013}),
  \urlprefix\url{http://pastel.archives-ouvertes.fr/pastel-00823758}.

\bibitem[{\citenamefont{de~Gironcoli}(1995)}]{deGironcoli:1995}
\bibinfo{author}{\bibfnamefont{S.}~\bibnamefont{de~Gironcoli}},
  \bibinfo{journal}{Phys. Rev. B} \textbf{\bibinfo{volume}{51}},
  \bibinfo{pages}{6773} (\bibinfo{year}{1995}).

\bibitem[{\citenamefont{Giannozzi et~al.}(2009)\citenamefont{Giannozzi, Baroni,
  Bonini, Calandra, Car, Cavazzoni, Ceresoli, Chiarotti, Cococcioni, Dabo
  et~al.}}]{Giannozzi:2009}
\bibinfo{author}{\bibfnamefont{P.}~\bibnamefont{Giannozzi}},
  \bibinfo{author}{\bibfnamefont{S.}~\bibnamefont{Baroni}},
  \bibinfo{author}{\bibfnamefont{N.}~\bibnamefont{Bonini}},
  \bibinfo{author}{\bibfnamefont{M.}~\bibnamefont{Calandra}},
  \bibinfo{author}{\bibfnamefont{R.}~\bibnamefont{Car}},
  \bibinfo{author}{\bibfnamefont{C.}~\bibnamefont{Cavazzoni}},
  \bibinfo{author}{\bibfnamefont{D.}~\bibnamefont{Ceresoli}},
  \bibinfo{author}{\bibfnamefont{G.}~\bibnamefont{Chiarotti}},
  \bibinfo{author}{\bibfnamefont{M.}~\bibnamefont{Cococcioni}},
  \bibinfo{author}{\bibfnamefont{I.}~\bibnamefont{Dabo}}, \bibnamefont{et~al.},
  \bibinfo{journal}{J. Phys.: Condens. Matter} \textbf{\bibinfo{volume}{21}},
  \bibinfo{pages}{395502} (\bibinfo{year}{2009}).

\bibitem[{\citenamefont{Perdew and Zunger}(1981)}]{Perdew:1981}
\bibinfo{author}{\bibfnamefont{J.}~\bibnamefont{Perdew}} \bibnamefont{and}
  \bibinfo{author}{\bibfnamefont{A.}~\bibnamefont{Zunger}},
  \bibinfo{journal}{Phys. Rev. B} \textbf{\bibinfo{volume}{23}},
  \bibinfo{pages}{5048} (\bibinfo{year}{1981}).

\bibitem[{Tim({\natexlab{b}})}]{Timrov:Note:2013:PP}
\bibinfo{note}{\url{http://www.quantum-espresso.org/pseudopotentials},
  \texttt{Si.pz-vbc.UPF} and \texttt{Al.pz-vbc.UPF}}.

\bibitem[{\citenamefont{Methfessel and Paxton}(1989)}]{Methfessel:1989}
\bibinfo{author}{\bibfnamefont{M.}~\bibnamefont{Methfessel}} \bibnamefont{and}
  \bibinfo{author}{\bibfnamefont{A.}~\bibnamefont{Paxton}},
  \bibinfo{journal}{Phys. Rev. B} \textbf{\bibinfo{volume}{40}},
  \bibinfo{pages}{3616} (\bibinfo{year}{1989}).

\bibitem[{\citenamefont{Neuberger}(1971)}]{Neuberger:1971}
\bibinfo{author}{\bibfnamefont{M.}~\bibnamefont{Neuberger}},
  \emph{\bibinfo{title}{Handbook of Electronic Materials}}
  (\bibinfo{publisher}{Plenum}, \bibinfo{address}{New York},
  \bibinfo{year}{1971}).

\bibitem[{\citenamefont{Wyckoff}(1963)}]{Wyckoff:1963}
\bibinfo{author}{\bibfnamefont{R.~W.} \bibnamefont{Wyckoff}},
  \emph{\bibinfo{title}{Crystal Structures}}, vol.~\bibinfo{volume}{1}
  (\bibinfo{publisher}{John Wiley and Sons Ltd}, \bibinfo{address}{New York},
  \bibinfo{year}{1963}), \bibinfo{edition}{2nd} ed.

\bibitem[{\citenamefont{Mahan}(1975)}]{Mahan:1990}
\bibinfo{author}{\bibfnamefont{G.~D.} \bibnamefont{Mahan}},
  \emph{\bibinfo{title}{Many-particles physics}} (\bibinfo{publisher}{Plenum
  Press}, \bibinfo{address}{New York}, \bibinfo{year}{1975}),
  \bibinfo{edition}{2nd} ed.

\bibitem[{Tim({\natexlab{c}})}]{Timrov:Note:2013:IXS}
\bibinfo{note}{The dynamic structure factor $S(\mathbf{Q},\omega)$ has been
  transformed to $-\mathrm{Im}[\epsilon^{-1}(\mathbf{Q},\mathbf{Q};\omega)]$
  using Eq.~(3) of Ref.~\onlinecite{Weissker:2010}, with the electron density
  being equal to 0.03 (a.u.)$^{-3}$.}

\bibitem[{Tim({\natexlab{d}})}]{Timrov:Note:2013:f-sum}
\bibinfo{note}{A remarkable feature of the Liouville-Lanczos approach is that
  the $f$-sum rule is satisfied exactly when truncating the Lanczos recursion
  to any number of iterations.\cite{Baroni:2012} The validity of this sum rule
  relies on the locality of the external potential. When non-local
  pseudopotentials are used, which is usually the case with plane-wave basis
  sets, violations of the $f$-sum rule occur,\cite{Malcioglu:2011} which we
  found to be sensitive to the accuracy of Brillouin zone sampling. In the
  present case we found these violations to be smaller than 7\%.}

\bibitem[{\citenamefont{Rocca et~al.}(2012)\citenamefont{Rocca, Ping, Gebauer,
  and Galli}}]{Rocca:2012}
\bibinfo{author}{\bibfnamefont{D.}~\bibnamefont{Rocca}},
  \bibinfo{author}{\bibfnamefont{Y.}~\bibnamefont{Ping}},
  \bibinfo{author}{\bibfnamefont{R.}~\bibnamefont{Gebauer}}, \bibnamefont{and}
  \bibinfo{author}{\bibfnamefont{G.}~\bibnamefont{Galli}},
  \bibinfo{journal}{Phys. Rev. B} \textbf{\bibinfo{volume}{85}},
  \bibinfo{pages}{045116} (\bibinfo{year}{2012}).

\end{thebibliography}
\end{document}